\documentclass[11pt]{article}

\usepackage{authblk}  
\usepackage{graphicx} 
\usepackage{epsfig}
\usepackage{dcolumn}  
\usepackage{bm}       
\usepackage{amsmath}
\usepackage{amssymb}
\usepackage{url}

\usepackage{float}

\usepackage{multirow}
\usepackage{booktabs}

\newcommand{\beq}{\begin{equation}} \newcommand{\eeq}{\end{equation}}
\newcommand{\bea}{\begin{eqnarray}} \newcommand{\eea}{\end{eqnarray}}

  \newcommand
{\Romannumeral}[1]{\uppercase\expandafter{\romannumeral#1}}

\newcommand{\be}{\begin{enumerate}} \newcommand{\ee}{\end{enumerate}}
\newcommand{\bi}{\begin{itemize}} \newcommand{\ei}{\end{itemize}}
\newcommand{\ba}{\begin{array}} \newcommand{\ea}{\end{array}}
\newcommand{\bc}{\begin{center}} \newcommand{\ec}{\end{center}}
\newcommand{\bt}{\begin{tabular}} \newcommand{\et}{\end{tabular}}

\def\lsim{\mathrel{\rlap{\lower4pt\hbox{\hskip1pt$\sim$}}
    \raise1pt\hbox{$<$}}}           
\def\gsim{\mathrel{\rlap{\lower4pt\hbox{\hskip1pt$\sim$}}
    \raise1pt\hbox{$>$}}}           

\newcommand{\Tr}{\mathop{\rm Tr}}           
\newcommand{\tr}{\mathop{\rm tr}}           
\newcommand{\half}{\textstyle {1\over2} \displaystyle}    
\newcommand{\third}{\textstyle {1\over3} \displaystyle}   
\newcommand{\quarter}{\textstyle {1\over4} \displaystyle} 
\newcommand{\eigth}{\textstyle {1\over8} \displaystyle}   
\newcommand{\twoth}{\textstyle {2\over3} \displaystyle}   
\newcommand{\Dslash}{{\hbox{D}\kern-0.6em\raise0.15ex\hbox{/}}} 

\renewcommand{\et}{\eta}

\begin{document}
	\thispagestyle{empty} 
	
	\setlength{\oddsidemargin}{0cm}
	\setlength{\baselineskip}{7mm}

\begin{normalsize}\begin{flushright}

October 2020 \\

\end{flushright}\end{normalsize}

\begin{center}
  
\vspace{15pt}

{\Large \bf Dyson's Equations for Quantum Gravity in the Hartree-Fock Approximation}

\vspace{30pt}

{\sl Herbert W. Hamber\footnote{HHamber@uci.edu.}, Lu Heng Sunny Yu \footnote{Lhyu1@uci.edu.}} 
\\
Department of Physics and Astronomy \\
University of California \\
Irvine, California 92697-4575, USA \\

\vspace{10pt}
\end{center}


\begin{center} 

{\bf ABSTRACT } 
\end{center}

Unlike scalar and gauge field theories in four dimensions, gravity is not perturbatively renormalizable and as a result perturbation theory is badly divergent.
Often the method of choice for investigating nonperturbative effects has been the lattice formulation, and in the case of gravity the Regge-Wheeler lattice path integral lends itself well for that purpose.
Nevertheless, lattice methods ultimately rely on extensive numerical calculations, 
leaving a desire for alternate methods that can be pursued analytically.
In this work we outline the Hartree-Fock approximation to quantum gravity, along lines
which are analogous to what is done for scalar fields and gauge theories.
The starting point is Dyson's equations, a closed set of integral equations which relate various physical amplitudes involving graviton propagators, vertex functions and proper self-energies.
Such equations are in general difficult to solve, and as a result not very useful in practice,
but nevertheless provide a basis for subsequent approximations.
This is where the Hartree-Fock approximation comes in, whereby lowest order diagrams get partially 
dressed by the use of fully interacting Green's function and self-energies, which then lead
to a set of self-consistent integral equations. 
The resulting nonlinear equations for the graviton self-energy show some 
remarkable features that clearly distinguish it from the scalar and gauge theory cases.
Specifically, for quantum gravity one finds a nontrivial ultraviolet fixed point in 
Newton's constant $G$ for spacetime dimensions greater than two, and
nontrivial scaling dimensions between $d=2$ and $d=4$, above which one obtains
Gaussian exponents.
In addition, the Hartree-Fock approximation gives an explicit analytic expression for the 
renormalization group running of Newton's constant, suggesting gravitational antiscreening 
with Newton's constant slowly increasing on cosmological scales.

\noindent

\newpage

\section{Introduction}
\label{sec:intro}  

\vskip 10pt

The traditional approach to quantum field theory is generally based on Feynman diagrams, which involve a perturbative expansion in some suitable small coupling constant. 
In cases like QED this procedure works rather well, since the vacuum about which one is expanding
is rather close, in its physical features, to the physical vacuum (weakly coupled electrons and photons).
In other cases, such as QCD, subtleties arise because of effects which are non-analytic in the coupling constant and cannot therefore be seen to any order in perturbation theory.
Indeed, in QCD it is well known that the perturbative ground state describes free quarks and gluons,
and as result the fundamental property of asymptotic freedom is easily derived in perturbation theory.
Nevertheless, important features such as gluon condensation, quark confinement and
chiral symmetry breaking remain invisible to any order in perturbation theory.
But the fact remains that the formulation of quantum field theory based on the Feynman path 
integral is generally not linked in any way to a perturbative expansion in terms of diagrams. 
As a result, the path integral generally makes sense even in a genuinely nonperturbative 
context, provided that it is properly defined and regulated, say via a spacetime lattice 
and a suitable Wick rotation.

Already at the level of nonrelativistic quantum mechanics, examples abound of physical system for which perturbation theory entirely fails to capture the essential properties of the model. 
Which leads to the reason why well-established nonperturbative methods such as the variational trial wave function method, numerical methods or the Hartree-Fock (HF) self-consistent method were developed early on in the history of quantum mechanics, motivated by the need to go beyond 
the - often uncertain - predictions of naive perturbation theory. 
In addition, many of these genuinely nonperturbative methods are relatively easy to implement, 
and in many cases lead to significantly improved physical predictions in atomic, molecular 
and nuclear physics.
Going one step further, examples abound also in relativistic quantum field theory and 
many-body theory, where naive perturbation theory clearly fails to give the correct answer. 
Interesting, and physically very relevant, cases of the latter include the theory of 
superconductivity, superfluidity, screening in a Coulomb gas, turbulent fluid flow, 
and generally many important aspects of critical (or cooperative) phenomena in statistical physics.
A common thread within many of these widely different systems is the emergence of 
coherent quantum mechanical behavior, which is not easily revealed by the use of perturbation 
theory about some non-interacting ground state.
However, some of the approximate methods mentioned above seem, at first, to be intimately tied 
to the existence of a Hamiltonian formulation (as in the case of the variational method), 
and are thus not readily available if one focuses on the Feynman path integral approach to 
quantum field theory.

Nevertheless, it is possible, in some cases, to cleverly modify perturbation theory in such a way 
as to make the diagrammatic perturbative approach still viable and useful. 
One such method was pioneered by Wilson, who suggested that in some theories a double 
expansion in both the coupling constant and the dimensionality (the so called $\epsilon$-expansion) could be performed [1-3].
These expansion methods turn out to be quite powerful, but they also generally 
lead to series that are known to be asymptotic, and in some cases (such as the nonlinear sigma 
model) can have rather poor convergence properties, mainly due to the existence of 
renormalon singularities along the positive real Borel transform axis.
Nevertheless, these investigations lead eventually to the deep insight that perturbatively 
nonrenormalizable theories might be renormalizable after all, if one is willing to go 
beyond the narrow context perturbation theory [4-6].
In addition, lately direct numerical methods have acquired a more prominent 
position, largely because of the increasing availability of very fast computers,
allowing many of the foregoing ideas to be rigorously tested.
A small subset of the vast literature on the subject of field theory methods combined with 
modern renormalization group ideas, as applied mostly to Euclidean quantum field theory
and statistical physics, can be found in [7-11].

Turning to the gravity context, it is not difficult to see that the case perhaps closest to 
quantum gravity is non-Abelian gauge theories, and specifically QCD. 
In the latter case one finds that, besides the established Feynman diagram perturbative approach in the 
gauge coupling $g$, there are very few additional approximate methods available, mostly due to the 
crucial need of preserving exact local gauge invariance. 
It is not surprising therefore that in recent years, for these theories, the method of choice has 
become the spacetime lattice formulation, which attempts to evaluate the Feynman path integral directly and exactly by numerical methods.
The above procedure then leads to answers which, at least in principle, can be improved arbitrarily 
given a fine enough lattice subdivision and sufficient amounts of computer time.
More recently lattice methods have also been applied to the case of quantum gravity, 
in the framework of the simplicial Regge-Wheeler discretization \cite{reg61,whe63,book}, where 
they have opened the door to accurate determinations of critical points and nontrivial scaling 
dimensions.

It would seem nevertheless desirable to be able to derive certain basic results in quantum gravity using perhaps approximate, but largely analytical methods. 
Unfortunately, in the case of gravity perturbation theory in Newton's $G$ is even less useful 
than for non-Abelian gauge theories and QCD, since the theory is unequivocally not 
perturbatively renormalizable in four spacetime dimensions [15-29].
To some extent it is possible to partially surmount this rather serious issue, by developing Wilson's 
$2+ \epsilon$ diagrammatic expansion \cite{wil73} for gravity, with $\epsilon = d-2 $.
Such an expansion is constructed around the two-dimensional theory, which, apart from being largely topological in nature, also leads in the end to the need to set 
$\epsilon = 2 $ (clearly not a small value) in order to finally reach the physical theory in $d=4$
\cite{gas78,chr78,wei79}.
Consequently, one finds significant quantitative uncertainties regarding the interpretation 
of the results in four spacetime dimensions, especially given
the yet unknown convergence properties of the asymptotic series in $\epsilon$.
Lately, approximate truncated renormalization group (RG) methods have been applied directly in four dimensions, which thus bypass the need for Wilson's dimensional expansion.
These nevertheless can lead to some significant uncertainties in trying to pin down the 
resulting truncation errors, in part due to what are often rather drastic renormalization 
group truncation procedures, which in turn generally severely limit the space of 
operators used in constructing the renormalization group flows.

In this work we will focus instead on the development of the Hartree-Fock approximation 
for quantum gravity. 
The Hartree-Fock method is of course well known in non-relativistic quantum mechanics and 
quantum chemistry, where it has been used for decades as a very successful tool for describing 
correctly (qualitatively, and often even quantitatively to a high degree) properties of atoms 
and molecules, including, among other things, the origin of the periodic table \cite{bet68}.
In the Hartree-Fock approach the direct interaction of one particle (say an electron in an atom) 
with all the other particles is represented by some sort of average field, where the average field itself 
is determined self-consistently via the solution of a nonlinear, but single-particle 
Schroedinger-like equation.
The latter in turn then determines, via the single-particle wave functions, the average field exerted on the original single particle introduced at the very start of the procedure.
As such, it is often referred to as a self-consistent method, where an explicit solution to the
effective nonlinear coupled wave equations is obtained by successive (and generally 
rapidly convergent) iterations.

It has been known for a while that the Hartree-Fock method has a clear extension to a quantum field 
theory (QFT) context.
In quantum field theory (and more generally in many-body theory) it is possible to derive the leading 
order Hartree-Fock approximation [33-44] as a truncation of the Dyson-Schwinger (DS) equations \cite{dys49,sch51,sch51a}.
For the latter a detailed, modern exposition can be found in the classic quantum field theory 
book \cite{itz80}, as well as (for non-relativistic applications) in \cite{fet71}.
The Hartree-Fock method is in fact quite popular to this day in many-body theory, where it is used, 
for example, as a way of  deriving the gap equation for a superconductor in the 
Bardeen-Cooper-Schrieffer (BCS) theory \cite{bcs57,bcs57a,bog58}.
There the Hartree-Fock method was originally used to show that the opening of a gap $\Delta$ 
at the Fermi surface is directly related to the existence of a Cooper pair condensate.
Many more applications of the Hartree-Fock method in statistical physics can be found in \cite{agd62,fet71}.
Later on, the successful use of the Hartree-Fock approach in many body theory spawned many useful 
applications to relativistic quantum field theory [39-44], including one of the earliest investigations
of dynamical chiral symmetry breaking.
There is also a general understanding that the Hartree-Fock method, since it is usually based on
the notion of a (self-consistently determined) average field, cannot properly account for the
regime where large fluctuations and correlations are important.
The precursor to these kind of reasonings is the general validity of mean field theory for spin systems
and ferromagnet above four dimensions (as described by the Landau theory), but not in low
dimensions where fluctuations become increasingly important.

In statistical field theory the Hartree-Fock approximation is seen to be closely related to the large-$N$ expansion, as described for example in detail in \cite{par81hf}. 
There one introduces $N$ copies for the fields in question (the simplest case is for a scalar field, but it can be applied to Fermions as well), and then expands the resulting Green's functions in 
powers of $1 / N $.
The expansion thus generally starts out from some sort of extended symmetry group theory, which is 
taken to be $O(N)$ or $SU(N)$ invariant.
Then it is easy to see that the leading order in the $1/ N $ expansion reproduces what one 
obtains to lowest order in the Hartree-Fock approximation for the same theory.
Furthermore, the next orders in the $1/ N $ expansion generally correspond to higher order 
corrections within the Hartree-Fock approximation.
The large $N$ expansion can be developed for $SU(N)$ gauge theories as well, and to
leading order it gives rise to the 't Hooft planar diagram expansion. 
In view of the previous discussion it would seem therefore rather natural to consider the 
analog of the Hartree-Fock approximation for QCD as well \cite{alk01,rob94}. 
Nevertheless, this avenue of inquiry has not been very popular lately, due mainly to its 
rather crude nature, and because of the increasing power (and much greater 
reliability) of ab initio numerical lattice approaches to QCD. 

Unfortunately, in the case of quantum gravity there is no obvious large $N$ expansion framework. 
This can be seen from the fact that one starts out with clearly only one spacetime metric, 
and introducing multiple $N$ copies of the same metric seems rather unnatural.
Such multiple metric formulation would cause some rather serious confusion 
on which of the $N$ metrics should be used to determine the spacetime interval between events,
and thus provide the needed spacetime arguments for gravitational $n$-point functions.
In addition, and unlike QCD, there is no natural global (or local) symmetry group that would
relate these different copies of the metric to each other. 
As will be shown below, it is nevertheless possible (and quite natural) to develop the Hartree-Fock
approximation for quantum gravity, in a way that is rather similar to the procedure followed 
in the case of the nonlinear sigma model, gauge theories and general many-body theories.
In the gravitational case one starts out, in close analogy to what is done for these other theories,
from the Dyson-Schwinger equations, and then truncates them to the lowest order terms, 
replacing the bare propagators by dressed ones, the bare vertices by dressed ones etc.
Thus, the Hartree-Fock approximation for quantum gravity can be written out without any 
reference to a $1 / N $ expansion, which as discussed above is often the underlying 
justification for other non-gravity theories.
In addition, the leading Hartree-Fock result represents just the lowest order term in a modified
diagrammatic expansion, which in principle can be carried out systematically to higher order.

As a reference, for the scalar $\lambda \phi^4 $ theory the Hartree-Fock approximation was already 
outlined some time ago in \cite{par81hf}, where its relationship to the $ 1/N $ expansion
was laid out in detail. 
One important aspect of both the Hartree-Fock approximation and the $ 1/N $ expansion discussed in the references given above is the fact that it can usually be applied in any dimensions, and is therefore not restricted in any way to the spacetime (or space) dimension in which the theory is found to
be perturbatively renormalizable.
It represents therefore a genuinely nonperturbative method, of potentially widespread application. 
In the scalar field case one can furthermore clearly establish, based on the comparison with 
other methods, to what extent the Hartree-Fock approximation is 
able to capture the key physical properties of the underlying theory, such as
correlation functions and universal scaling exponents.
These investigations in turn provide a partial insight into what physical aspects the Hartree-Fock approximation can correctly reproduce, and where it fails.
Generally, the conclusion appears to be that, while quantitatively not exceedingly accurate, the 
Hartree-Fock approximation and the $ 1/ N$ expansion tend to reproduce correctly at least
the qualitative features of the theory in various dimensions, including, for example, the existence 
of an upper and a lower critical dimension, the appearance of nontrivial fixed points of the
renormalization group, as well as the dependence of nontrivial scaling exponents on the 
dimension of spacetime.

In this work we first review the properties of the Hartree-Fock approximation, as it applies initially to some of the simpler field theories. 
In the case of the scalar field theory, with the scalar field having $N$ components, one can see that 
the Hartree-Fock approximation is easily obtained from the large $N$ saddle point \cite{par81hf}. 
The resulting gap equation is generally valid in any dimension, and once solved provides an 
explicit expression for the mass gap (the inverse correlation length) as a function of the bare 
coupling constant.
Solutions to the gap equation show some significant sensitivity to the number of dimensions. 
For the scalar field a clear transition is seen between the lower critical dimension $d=2$ (for $N>2$)
and the upper critical dimension $d=4$, above which the critical exponents attain the Landau
theory values. 
Indeed, the Hartree-Fock approximation reproduces correctly the fact that for $ d \le 2 $ the $O(N)$ 
symmetric nonlinear sigma model for $N>2$ exhibits a non-vanishing gap for bare coupling $ g > 0 $, 
and that for $ d \ge 4 $  it reduces to a non-interacting (Gaussian) theory in the long-distance limit 
(also known as the triviality of $\lambda \, \phi^4 $ in $ d \ge 4 $).
Then, as a warm-up to the gravity case, it will pay here to discuss briefly the $SU(N)$ gauge theory case.
Here again one can see that the Hartree-Fock result correctly reproduces the asymptotic freedom 
running of the gauge coupling $g$, and furthermore shows that $d=4$ is the lower critical dimension
for $SU(N)$ gauge theories. 
This last statement is meant to refer to the fact that above four dimensions non-Abelian gauge 
theories are known to have a nontrivial ultraviolet fixed point at some $g_c$, separating 
a weak coupling Coulomb-like phase from a strong coupling confining phase.

Finally, in the gravitational case one starts out by following a procedure which closely 
parallels the gauge theory case.
Nevertheless, at the same time one is faced with the fact that the gravitational case is considerably 
more difficult, for a number of (mostly well known) reasons which we proceed to enumerate here.
Unlike the scalar field and gauge theory case in four dimensions, gravity is not perturbatively renormalizable in four dimensions, and as a result perturbation theory is badly divergent, and thus
not very useful.
Furthermore, as mentioned previously, there is no sensible way of formulating a large $N$
expansion for gravity.
That leaves as the method of choice, at least for investigating nonperturbative properties of the theory,
the lattice formulation, and specifically the elegant simplicial lattice formulation due to Regge and 
Wheeler \cite{reg61,whe63} and it's Euclidean Feynman path integral extension.
One disadvantage is that the lattice methods nevertheless ultimately rely on extensive
numerical calculations, leaving a desire for alternate methods that can be pursued analytically.
As mentioned previously, it is also possible to develop for quantum gravity the 
$2+\epsilon$ dimensional expansion \cite{gas78,chr78,wei79},
in close analogy to Wilson's $4-\epsilon$ and $2+\epsilon$ dimensional expansion 
for scalar field theories \cite{wil73};
the ability to perform such an expansion for gravity hinges on the fact that Einstein's theory 
becomes formally perturbatively renormalizable in two dimensions.
But there are significant problems associated with this expansion, notably the fact 
that $\epsilon = d-2$ has to be set equal to two at the end of the calculation.
One more possible approach is via the Wheeler-DeWitt equations, which provide a useful Hamiltonian
formulation for gravity, both in the continuum and on the lattice, see \cite{hw11} and references
therein.
Nevertheless, while it has been possible to obtain some exact results on the lattice in $2+1$ 
dimensions, the $3+1$ case still appears rather difficult.
As a result, generally the number of reasonably well-tested nonperturbative methods available to 
investigate the vacuum structure of quantum gravity seem rather limited.
It is therefore quite remarkable that it is possible to develop a Hartree-Fock
approximation to quantum gravity, in a way that in the end is completely analogous to what is 
done in scalar field theories and gauge theories.
In the following we briefly outline the features of the Hartree-Fock method as applied to gravity.

For any field theory it is possible to write down a closed set of integral equations, which relate various physical amplitudes involving particle propagators, vertex functions, proper self-energies etc.
These equations follow in a rather direct way from their respective definitions in terms of the Feynman path integral, and the closely connected generating function $Z[J]$.
One particularly elegant and economic way to obtain these expressions is to apply suitable 
functional derivatives to the generating function in the presence of external classical sources.
The resulting exact relationship between $n$-point functions, known as the set of Dyson-Schwinger
equations, are, by virtue of their derivation, not reliant in any way on perturbation theory,
and thus genuinely nonperturbative.
Nevertheless, the set of coupled Dyson-Schwinger equations for the propagators, self-energies and vertex functions are in general very difficult to solve, and as a result not very useful in practice.
This is where the Hartree-Fock approximation comes in, whereby lowest order diagrams get partially 
dressed by the use of fully interacting Green's function and self-energies.
Since the resulting coupled equations involve fully dressed propagators and full proper self-energies
to some extent, they are referred to as a set of self-consistent equations. 
In principle the Hartree-Fock approximation to the Dyson-Schwinger equations can be carried out to 
arbitrarily high order, by examining the contribution of increasingly complicated diagrams involving dressed propagators and dressed self-energies etc.
Of particular interest here will be therefore the lowest order Hartree-Fock approximation for quantum gravity.
As in the scalar and gauge theory case, it will be necessary, in order to derive the Hartree-Fock equations, to write out the lowest order contributions to the graviton propagator, to the gravitational self-energy, and
to the gravitational vertex function, and for which the lowest order perturbation theory expressions 
are known as well.
Indeed, one remarkable aspect of most Hartree-Fock approximations lies in the fact that ultimately 
the calculations can be shown to reduce to the evaluation of a single one-loop integral, 
involving a self-energy contribution, which is then determined self-consistently.

Nevertheless, quite generally the resulting diagrammatic expressions are still ultraviolet divergent, 
and the procedure thus requires the introduction of an ultraviolet regularization, such as the one 
provided by dimensional regularization, an explicit momentum cutoff, or a lattice as in 
the Regge-Wheeler lattice formulation of gravity \cite{reg61,whe63}.
Once this is done, one then finds a number of significant similarities of the Hartree-Fock
result with the (equally perturbatively non-renormalizable in $d>2$) nonlinear sigma model.
Specifically, in the quantum gravity case one finds that the ultraviolet fixed point in 
Newton's constant $G$ is at $G=0$ in two spacetime dimensions, 
and moves to a non-zero value above $d=2$.
Nontrivial scaling dimensions are found between $d=2$ and $d=4$, above which one obtains
Gaussian (free field) exponents, thus suggesting that, at least in the Hartree-Fock approximation, 
the upper critical dimension for gravity is $d=4$.
One useful and interesting aspect of the above calculations is that the Hartree-Fock results are 
not too far off from what one obtains by either numerical investigations within the 
Regge-Wheeler lattice formulation, or via the $2+\epsilon$ dimensional expansion, 
or by truncated renormalization group methods.
Furthermore, as will be shown below, the Hartree-Fock approximation provides an explicit expression for the renormalization group (RG) running of Newton's constant, just like the Hartree-Fock leads to similar results for the nonlinear sigma model.
As in the lattice case, the Hartree-Fock results suggest a two-phase structure, with gravitational screening for $G < Gc$, and gravitational antiscreening for $G > Gc$ in four dimensions.
Due to the known nonperturbative instability of the weak coupling phase of $G<G_c$, only
the $G > G_c$ phase will need to be considered further, which then leads to
a gravitational coupling $ G (k) $ that slowly increases with distance, with a momentum
dependence that can be obtained explicitly within the Hartree-Fock approximation.

The paper is organized as follows. 
In Section 2 we recall the main features of Dyson's equations and how they lead to the 
Hartree-Fock approximation.
We point out the important and general result that Dyson's equations can be easily derived using
the elegant machinery of the Feynman path integral combined with functional methods.
The Hartree-Fock approximation then follows from considering a specific set of sub-diagrams
with suitably dressed propagators, all to be determined later via the solution of a set of
self-consistent equations.
In Section 3 we discuss in detail, as a first application and as preamble to the quantum gravity case,
the case of the $O(N)$-symmetric nonlinear sigma model, initially here from the perturbative 
perspective of Wilson's $2+\epsilon$ expansion.
A summary of the basic results will turn out to be useful when comparing later to the Hartree-Fock
approximation.
In Section 4 we briefly recall the main features of the nonlinear sigma model in the large-$N$ expansion,
which can be carried out in any dimension.
Again, the discussion here is with an eye towards the later, more complex discussion for gravity.
In Section 5 we discuss the main features of the Hartree-Fock result for the $O(N)$ nonlinear
sigma model in general dimensions $ 2 \leq d \leq 4 $, and
point out its close relationship to the large-$N$ results described earlier.
The section ends with a comparison of the three approximations, and how they favorably 
relate to each other and to other results, such as the lattice formulation and actual experiments.
In Section 6 we proceed to a more difficult case, namely the discussion of some basic results for
the Hartree-Fock approximation for non-Abelian gauge theories in general spacetime 
dimensions $ 4 \leq d \leq 6 $.
The basic starting point is again a general gap equation for a nonperturbative, dynamically
generated mass scale.
Here again we will note that the Hartree-Fock results capture many of the basic ingredients of
nonperturbative physics, while still missing out on some (such as confinement).
In Section 7 we move on to the gravitational case, and discuss in detail the results of the Hartree-Fock
approximation in general spacetime dimensions $ 2 \leq d \leq 4 $.
For the gravity case, the basic starting point will be again a gap equation for the nonperturbatively
generated mass scale, whose solution will lead to a number of explicit results, including an explicit
expression for the renormalization group running of Newton's constant $G$.
In Section 8 we then present a sample application of the Hartree-Fock results to cosmology, 
where the running of $G$ is compared
to current cosmological data, specifically here the Planck-18 Cosmic Background Radiation (CMB) data.
Finally, Section 9 summarizes the key points of our study and concludes the paper.


\vskip 20pt

\section{Dyson's Equations and the Hartree-Fock Approximation}
\label{sec:hartree} 

\vskip 10pt

One of the earliest applications of the Hartree-Fock approximation
to solving Dyson's equations for propagators and vertex functions
was in the context of the BCS theory for superconductors \cite{bcs57,bcs57a,bog58}.
A few years later it was applied to the (perturbatively nonrenormalizable) 
relativistic theory of a self-coupled Fermion, where it provided the 
first convincing evidence for a dynamical breaking of chiral symmetry and the
emergence of Nambu-Goldstone bosons \cite{njl61,njl61a}.
A necessary preliminary step involved in deriving the Hartee-Fock approximation to a given
theory is writing down Dyson's equations (often referred to as the Schwinger-Dyson 
equations) for the field, or fields, in question.
Dyson's equations have a nice, almost self-explanatory, form when written in terms of Feynman
diagrams for various propagators, self-energies and vertex functions associated with the 
particles appearing in the theory, such as for example electrons and photons in QED.
These generally describe a set of coupled integral equations for the various dressed,
fully interacting, $n$-point functions, but whose solution is generally rather difficult,
if not impossible.
As a result, they are usually only discussed as a method for deriving exact Ward identities
that follow from local gauge invariance, with their practical application then usually
restricted in most cases to perturbation theory only.
Nevertheless, as will be shown below, they play a key role in deriving the Hartree-Fock
approximations for the fields in question, thus providing a method that is essentially
nonperturbative, involving a partial re-summation of an infinite class of diagrams.


In non-relativistic many body theory Dyson's equation take on a simple form when
describing the static electromagnetic interaction between electrons \cite{fet71}.
\footnote{
It is often customary in many body theory to describe the interaction of electrons 
close to the Fermi surface in terms of electrons and holes situated around the conduction band, 
in analogy to the treatment of relativistic electrons in the framework of the Dirac equation,
with the positively charged holes playing the role of positrons in the Dirac theory.}
In terms of the electron propagator $G(p)$ one has at first the decomposition
\beq
G(p) \; = \; G_0 (p) \, + \, G_0 (p) \, \Sigma (p) \, G_0 (p) \;\; ,
\eeq
where $G_0 (p)$ is the bare (tree level) propagator, $G(p)$ the dressed one and 
$\Sigma (p) $ the electron self-energy.
The above should really be regarded as a matrix equation, with indices appropriate
for the particle being exchanged intended, and not written out in this case.
If one regards the self-energy $\Sigma $ as being constructed out of repeated
iterations of a proper self-energy $\Sigma^{*} (p) $, then Dyson's equation
takes on the form
\beq
G(p) \; = \; G_0 (p) \, + \, G_0 (p) \, \Sigma^{*}  (p) \, G (p)
\eeq
with solution
\beq
G(p) \; = \; { 1 \over G_0^{-1}  (p) \, - \, \Sigma^{*}  (p) }  \;\;. 
\eeq
Since generally $G(p) $ is regarded as a matrix with spin indices, the above should be
interpreted as the matrix inverse.
This then leads to the identification of $\Sigma^{*} (0) $ as an
effective mass correction for the electron, induced, perturbatively or nonperturbatively,
by radiative corrections.

Similarly, the boson (photon or phonon, depending of the interaction being considered) exchange between electrons leads to the expansion for the interaction term
\beq
U (p) \; = \; U_0 (p) \, + \, U_0 (p) \, \Pi (p) \, U_0 (p)
\eeq
where $U_0 (p)$ is the bare (tree level) contribution, $U (p)$ the dressed one and 
$\Pi (p) $ the self-energy contribution.
Again, the above should be regarded as a matrix equation, with indices appropriate
for the particle being exchanged.
If one considers the above self-energy $\Pi (p) $ as being constructed out of repeated
iterations of a proper self-energy $\Pi^{*} (p) $, then Dyson's equation
in this case takes on the form
\beq
U (p) \; = \; U_0 (p) \, + \, U_0 (p) \, \Pi^{*}  (p) \, U (p)
\eeq
with solution
\beq
U (p) \; = \; { 1 \over U_0^{-1}  (p) \, - \, \Pi^{*}  (p) }  
\eeq
again intended as a proper matrix inverse.
If the interaction is spin-independent, as in the Coulomb case, then
the above matrices can all be regarded as diagonal.
This last result then leads to the identification of $\Pi^{*} (0) $ as the 
effective mass (or inverse range) associated with the interaction described by $U(p)$.
The ratio $ \kappa (p) = U_0 (p) / U(p) $ is referred to, in the electromagnetic
case, as an effective generalized dielectric function \cite{fet71}.


Alternatively, it is possible to derive Dyson's equations directly from the Feynman path
integral using the rather elegant machinery of functional differentiation of the generating
function $Z[J]$ with respect to a suitable source terms $J(x)$.
It is this latter approach that we follow here, as it generalizes rather straightforwardly to 
the case of quantum gravity.
For a scalar field one has the statement that the integral of a derivative is zero for suitable boundary conditions, or 
\beq
\int [d \, \phi ] \, {\delta \over \delta \phi (x) } \, 
e^{ i \int dx \left ( \mathcal{L} + J \, \phi \right ) } \; = \; 0
\eeq
where $\mathcal{L} $ is the Lagrangian density for the scalar field in question.
As a consequence
\beq
\int [ d \, \phi ] \, 
\left [ {\partial \, \mathcal{L} \over \partial \, \phi (x) } \, + \, J(x) \right ]
e^{ i \int dx \left ( \mathcal{L} + J \, \phi \right ) } \; = \; 0  \; .
\eeq
If the scalar field potential is generally denoted by $V(\phi) $, then the previous expression can
be re-cast, defining
\beq
Z[J] \; =\; \int [ d \, \phi ] \, e^{ i \int dx \left ( \mathcal{L} + J \, \phi \right ) }
\eeq
as
\beq
\left [ \, ( \Box_x + m^2 ) \, { \delta \over i \, \delta J (x) } \, + \, 
V' \left  ( { \delta \over i \, \delta J (x) }  \right ) \, - \, J(x) \right ] \;\; 
Z[J] \; =\; 0
\eeq
The latter then plays the role of the quantum Heisenberg equations of motion.
Successive applications of functional derivatives with respect to $J(x)$ then
leads to a sequence of simple identities at zero source
\beq
\int [ d \, \phi ] \, 
\left [ \, ( \Box_x + m^2 ) \, \phi (x)  \, + \, V' ( \phi(x) ) \, \right ] \, 
e^{ i \int dx \, \mathcal{L} }
\;  = \; 0
\eeq
\beq
\int [ d \, \phi ] \, 
\left [ ( \Box_x + m^2 ) \, \phi (x)  \cdot \phi (x_1 ) \, + \, V' ( \phi(x) ) \cdot  \phi (x_1 ) \right ] \, 
e^{ i \int dx \, \mathcal{L} }
\;  = \; 0
\eeq
and so on.
More generally, for $n$-point functions one obtains
\bea
& \int [ d \, \phi ] \,  &
\left [ ( \Box_x + m^2 ) \phi (x) \, + \, V' ( \phi(x) ) \right ] \, \phi (x_1) \, \phi (x_2 ) \dots \phi ( x_n) \;
e^{ i \int dx \, \mathcal{L} }
\nonumber \\
& + & 
\int [ d \, \phi ] \,  \delta ( x - x_1 ) \, \phi (x_2) \dots \phi ( x_n) \;
e^{ i \int dx \, \mathcal{L} }
\nonumber \\
& + & \dots + 
\int [ d \, \phi ] \, \phi (x_1) \, \phi (x_2) \dots \delta ( x - x_n ) \;
e^{ i \int dx \, \mathcal{L} }
\;  = \; 0
\eea

\begin{figure}
	\begin{center}
		\includegraphics[width=0.75\textwidth]{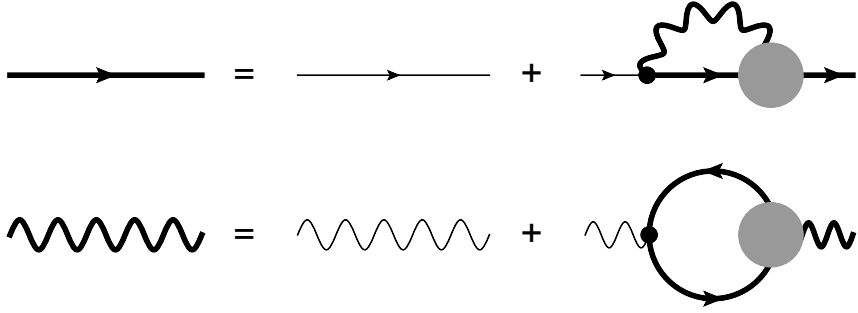}
	\end{center}
	\caption{Dyson's equations in QED relating the dressed fermion propagator, the 
               dressed photon propagator and the dressed vertex function. Thin lines represent 
       bare propagators, while thicker lines represent the dressed ones.}	
	\label{fig:qed_dyson}
\end{figure}

Within the current discussion of Dyson's equations, a step up in difficulty is clearly represented by
the gauge theory case, with QED as the simplest example.
Since the resulting equations share some similarity with quantum gravity, it will be useful here to
recall briefly the main aspects.
Compared to scalar field theory, the first new ingredient is the presence of multiple fields,
for QED in the form of photon and matter fields.
The generating function is
\beq
Z ( J, \eta, {\bar \eta} ) \; = \; e^{ G ( J, \eta, {\bar \eta} ) } \; = \;
\int [ d A] [ d \psi ] [ d {\bar \psi} ] \; e^{ i S }
\eeq
with
\beq
S \; = \; I \, + \, \int dx \left ( J \cdot A  + {\bar \eta} \, \psi + {\bar \psi } \, \eta \right )
\eeq
where $ I ( A, \psi , {\bar \psi } ) $ is the QED action.
Again, for suitable boundary conditions the integral of a derivative is zero, which leads to
the identity
\beq
\int [ d A] [ d \psi ] [ d {\bar \psi} ] \; { \delta \over \delta A^\mu (x) } \; e^{ i S } \; =\; 0 \;\; ,
\label{eq:qed_id}
\eeq
and consequently
\beq
\int [ d A] [ d \psi ] [ d {\bar \psi} ] \, 
\left ( { \delta\,  I \over \delta A^\mu (x) } \, + \, J_\mu \right ) \,  e^{ i S } \; =\; 0 \;\; .
\label{eq:qed_id1}
\eeq
For the case of the QED action one has
\beq
{ \delta\,  I \over \delta A^\mu (x) } \; = \; 
\left [ \, \Box g_{\mu\nu} - (1-\lambda) \partial_\mu \partial_\nu \, \right ] A^\nu  
\, - \, e \, {\bar \psi} \gamma_\mu \, \psi  \;\; ,
\label{eq:qed_id2}
\eeq
where $\lambda$ is the gauge parameter for a gauge fixing term $ \half \lambda (\partial \cdot A )^2 $,
and $g_{\mu\nu}$  here the flat Lorentz metric.
The resulting equation 
\beq
J_\mu \, + \,  
\left [ \, \Box g_{\mu\nu} - (1-\lambda) \partial_\mu \partial_\nu \, \right ] 
{ \delta G \over i \, \delta J^\nu  } \, - \, 
e \, { \delta \, G \over \delta \, \eta } \, \gamma_\mu \, 
{ \delta \, G \over \delta \, {\bar \eta } }
\, - \,  
e \, { \delta \over \delta \, \eta } \, 
\left ( \gamma_\mu \, { \delta \, G \over \delta \, {\bar \eta }  } \right )
\; = \; 0
\label{eq:max_quant}
\eeq
can then be regarded as the quantum version of Maxwell's equations \cite{itz80}.
A graphical illustration of Dyson's equation for QED is given in Figure \ref{fig:qed_dyson}.
In QED and gauge theories in general it is often useful to introduce the generating
function $\Gamma$ of one-particle-irreducible (1PI) Green's functions, 
defined as the Legendre transform of $G$, 
\beq
G ( J, \eta, {\bar \eta} )  \, = \, i \, \Gamma ( A, \psi , {\bar \psi } ) \, + \, 
i \int dx \, \left ( J \cdot A  + {\bar \eta} \, \psi + {\bar \psi } \, \eta \right )  \;\; .
\eeq
Then the introduction of the classical fields $A_\mu$, $\psi $ and ${\bar \psi} $ via the definitions
\beq
{ \delta \, G \over i \delta \, J^\mu }  \, = \, A_\mu 
\;\;\;\; \;\;\;\;  
- { \delta \, G \over i \delta \, \eta }  \, = \, {\bar \psi }
\;\;\;\; \;\;\;\;  
{ \delta \, G \over i \delta \, {\bar \eta }  } \, = \, \psi
\eeq
and conversely, as a consequence of the definition of $\Gamma$,
\beq
{ \delta \, \Gamma \over \delta \, A^\mu }  \, = \, - \, J_\mu 
\;\;\;\; \;\;\;\;  
{ \delta \, \Gamma \over \delta \, \psi }  \, = \, {\bar \eta }
\;\;\;\; \;\;\;\;  
{ \delta \, \Gamma \over \delta \, {\bar \psi }  } \, = \, - \, \eta
\eeq
which allows one to rewrite Eq.~(\ref{eq:max_quant}) simply as 
\beq
J_\mu \, + \,  
\left [ \, \Box g_{\mu\nu} - (1-\lambda) \partial_\mu \partial_\nu \, \right ] \, A^\nu
\, - \, e \, {\bar \psi } \, \gamma_\mu \, \psi 
\, - \,  i e \, { \delta \over \delta \, \eta } \, \left ( \gamma_\mu \, \psi  \right ) \; = \; 0 \;\; .
\label{eq:max_quant1}
\eeq
Equivalently, one has in terms of the generating function $\Gamma $ exclusively
\beq
{ \delta \Gamma \over \delta \, A^\mu } 
\; = \; 
\left [ \, \Box g_{\mu\nu} - (1-\lambda) \partial_\mu \partial_\nu \, \right ] \, A^\nu
\, - \, e \, {\bar \psi } \, \gamma_\mu \, \psi \, + \,  
i e \, \Tr  
\left [ \gamma_\mu 
\left ( { \delta^2 \, \Gamma \over \delta \, \psi \, \delta {\bar \psi } } \right )^{-1} 
\right ]  \;\; .
\label{eq:max_quant2}
\eeq


A very similar procedure can be carried out for non-Abelian gauge theories and gravity, 
nevertheless the complexity increases greatly due to the proliferation of
Lorentz indices, color indices and a multitude of interaction vertices.
We will skip here almost entirely a detailed discussion of gauge theories, and only refer to the
relevant diagrams and corresponding expressions, where appropriate.
Also, for ease of exposition, the quantum gravity case will be postponed until the Hartree-Fock
approximation for gravity is developed.


\vskip 20pt

\section{The Case of the Nonlinear Sigma Model}
\label{sec:sigma_pert}  

\vskip 10pt

The question then arises, are there any field theories where the standard perturbative treatment fails, yet for which one can find alternative methods and from them develop consistent predictions?
The answer seems unequivocally yes \cite{par73,par75,par76}.
Indeed, outside the quantum gravity framework, there are two notable examples of field theories: the nonlinear sigma model and the self-coupled fermion model. 
Both are not perturbatively renormalizable for $d>2$, and yet lead to
consistent, and in some instances precision-testable, predictions above $d=2$.
The key ingredient to all of these results is, as recognized originally by Wilson, the existence of a nontrivial ultraviolet fixed point (also known as a phase transition in the statistical field theory context)
with nontrivial universal scaling dimensions \cite{wil72,wil72a,wil73,gro74}.
Furthermore, several different calculational approaches are by now available
for comparing predictions: the $2+\epsilon$ expansion, the large-$N$
limit, the functional renormalization group, the conformal bootstrap, and a variety of
lattice approaches.
From within the lattice approach, several additional techniques are available:
the Wilson (block-spin) renormalization group approach, the strong coupling expansion, 
the weak coupling expansion and the numerically exact evaluation of the lattice path integral.

The $O(N)$-symmetric nonlinear sigma model provides an instructive (and
rich) example of a theory which, above two dimensions, is not 
perturbatively renormalizable in the traditional sense, and
yet can be studied in a controlled way via Wilson's $2+\epsilon$ expansion [53-60].
Such a framework provides a consistent way to calculate nontrivial
scaling properties
of the theory in those dimensions where it is not perturbatively
renormalizable (e.g. $d=3$ and $d=4$), which can then be compared
to nonperturbative
results based on the lattice theory.
In addition, the model can be solved exactly in the large $N$ limit for 
any $d$, without any reliance on the $2+\epsilon$ expansion.
In all three approaches the model exhibits a nontrivial
ultraviolet fixed point at some coupling $g_c$ (a phase transition
in statistical mechanics language), separating a weak coupling massless
ordered phase from a massive strong coupling phase.
Finally, the results can then be compared to experiments,
since in $d=3$ the model describes either a ferromagnet or superfluid helium
in the vicinity of its critical point.


The nonlinear sigma model is described by an $N$-component scalar field $\phi_a$
satisfying a unit constraint $\phi^2(x)=1$, with functional integral given by
\bea
Z [J] \, & = & \, \int [ \, d \phi \, ] \, 
\prod_x \, \delta \left [ \phi (x) \cdot \phi (x) - 1 \right ] \,
\nonumber \\
&& \times \,
\exp \left ( - \, { \Lambda^{d-2} \over g^2 } \, S(\phi) 
\, + \int d^d x \; J(x) \cdot \phi(x) \, \right ) \;\; .
\label{eq:sigma_cont}
\eea
The action is taken to be $O(N)$-invariant
\beq
S(\phi) \, = \, \half \,
\int d^d x \; \partial_\mu \phi (x) \cdot \partial_\mu \phi (x)  \;\; .
\eeq
Here $\Lambda$ is an ultraviolet cutoff (as provided for example by a lattice),
and $g$ the bare dimensionless coupling at the cutoff scale $\Lambda$;
in a statistical field theory context the coupling $g^2$ plays the role of a temperature.

In perturbation theory one can eliminate one $\phi$ field by 
introducing a convenient parametrization for the unit sphere,
$\phi (x) = \{ \sigma(x), \mathbf{\pi}(x) \}$ where $\pi_a$ is an
$N-1$-component field, and then solving locally for $\sigma(x)$
\beq
\sigma (x) \, = \, [ \, 1 - \mathbf{\pi}^2 (x) \, ]^{1/2}  \;\; .
\eeq
In the framework of perturbation theory in $g$ 
the constraint $ | \mathbf{\pi} (x) | < 1$
is not important as one is restricting the fluctuations
to be small.
Accordingly the $\mathbf{\pi}$ integrations are extended from
$- \infty$ to $+\infty$, which reduces the development
of the perturbative expansion to a sequence of Gaussian integrals.
Values of $\mathbf{\pi} (x) \sim 1 $ give exponentially small
contributions of order $\exp (- {\rm const.} / g ) $ which are
considered negligible to any finite order in perturbation theory.
In term of the $\mathbf{\pi}$ field the original action $S$ becomes
\beq
S( \mathbf{\pi}) \, = \, \half \,
\int d^d x \, \left [ \,
( \partial_\mu \mathbf{\pi} )^2 
+ { ( \mathbf{\pi} \cdot \partial_\mu \mathbf{\pi} )^2 \over 1 
- \mathbf{\pi}^2 } \, \right ]  \;\; .
\eeq
The change of variables from $\phi(x)$ to $\mathbf{\pi}(x) $ gives rise to a Jacobian
\beq
\prod_x \left [ \, 1 - \mathbf{\pi}^2 \, \right ]^{-1/2} \, \sim \;
\exp \left [ \, - \half \, \delta^d (0) \int d^d x \,
\log ( 1 - \mathbf{\pi}^2 ) \, \right ] \;\; ,
\label{eq:jacobian}
\eeq
which is necessary for the cancellation of spurious tadpole divergences.
In the presence of an explicit ultraviolet cutoff $\delta (0) \simeq ( \Lambda / \pi )^d $.
The combined functional integral for the unconstrained
$\mathbf{\pi}$ field is then given by
\beq
Z [ J ] = \int [ \, d \mathbf{\pi} \, ] \, 
\exp \left ( - \, { \Lambda^{d-2} \over g^2 } \, S_{0}( \mathbf{\pi} ) 
\, + \int d^d x \; J (x) \cdot \mathbf{\pi} (x) \right )
\label{eq:sigma_cont1}
\eeq
with 
\beq
S_{0}( \mathbf{\pi} ) \, = \, \half \int d^d x \, 
\left [ 
( \partial_\mu \mathbf{\pi} )^2 
+ { ( \mathbf{\pi} \cdot \partial_\mu \mathbf{\pi} )^2 \over 1 - \mathbf{\pi}^2 }
\right ] 
\, + \, \half \, \delta^d (0) \int d^d x \; \log ( 1 - \mathbf{\pi}^2 )  \;\; .
\eeq
In perturbation theory the above action is then expanded out in powers
of $\mathbf{\pi}$, and the propagator for the $\mathbf{\pi}$ field can be read off
from the quadratic part of the action,
\beq
\Delta_{ab} (k^2) \, = \, { \delta_{ab} \over k^2 }  \;\; .
\eeq
In the weak coupling limit the $\mathbf{\pi}$ fields correspond
to the Goldstone modes of the spontaneously broken $O(N)$
symmetry, the latter broken spontaneously in the ordered phase by
a non-vanishing vacuum expectation value $\langle \mathbf{\pi} \rangle \neq 0$.
Since the $\mathbf{\pi}$ field has mass dimension $\half (d-2)$, and thus the
interaction $\partial^2 \mathbf{\pi}^{2n}$ consequently has dimension
$n (d-2) +2 $, one finds that the theory is perturbatively renormalizable in
$d=2$, and perturbatively non-renormalizable above $d=2$.
Potential infrared problems due to massless propagators
are handled by introducing an
external $h$-field term for the original composite $\sigma$ field,
which then can be seen to act as a regulating mass term for the $\mathbf{\pi}$ field.

One can write down the same field theory on a lattice, where it
corresponds to the $O(N)$-symmetric classical Heisenberg
model at a finite temperature $ T \sim g^2 $.
The simplest procedure is to introduce a hypercubic lattice
of spacing $a$, with sites labeled by integers 
$\mathbf{n}=(n_1 \dots n_d)$,
which introduces an ultraviolet cutoff $\Lambda \sim \pi/a$.
On the lattice, field derivatives are replaced by finite differences
\beq
\partial_\mu \phi (x) \; \rightarrow \; 
\Delta_\mu \phi ( {\bf n} ) \, \equiv \,
{ \phi ( {\bf n} + {\bf \mu} ) - \phi ( {\bf n} ) \over a }
\eeq
and the discretized path integral then reads
\bea
Z[\, J \, ] & = & \, \int \, \prod_{\bf n} d \phi ( {\bf n} )
\, \delta [ \phi^2 ( {\bf n} ) -1 ] \,
\nonumber \\
& \times & \exp \left [ - {a^{2-d} \over 2 g^2 } \sum_{\bf n, \mu}
\left ( \Delta_\mu \phi ( {\bf n} ) \right )^2 +
\sum_{\bf n} J( {\bf n} ) \cdot \phi ( {\bf n} ) \right ]   \;\; .
\label{eq:sigma_lattice}
\eea
The above expression is recognized as the partition
function for a ferromagnetic $O(N)$-symmetric lattice spin system at finite temperature. 
Besides ferromagnets, it can be used to describe systems which
are related to it by universality, such as superconductors and
superfluid helium transitions.

In two dimensions one can compute the renormalization of the coupling
$g$ from the action of Eq.~(\ref{eq:sigma_cont1}) and one finds
after a short calculation \cite{pol75,bre76} for small $g$
\beq
{ 1 \over g^2 (\mu) } \; = \; { 1 \over g^2} \, 
+ \, { N - 2 \over 8 \pi } \, \log { \mu^2 \over \Lambda^2 } 
\, + \, \dots
\eeq
where $\mu$ is an arbitrary momentum scale.
Physically one can view the origin of the factor of $N-2$ in the fact that
there are $N-2$ directions in which the spin can experience
rapid small fluctuations perpendicular to its average
slow motion on the unit sphere, and that only these fluctuations
contribute to leading order.
In two dimensions the quantum correction (the second term on the r.h.s.)
increases the value of the effective coupling at low momenta (large
distances), unless $N=2$ in which case the correction vanishes.
There the quantum correction can be shown to vanish to all orders
in this case, nevertheless the vanishing of the $\beta$-function in two dimensions
for the $O(2)$ model is true only in perturbation theory.
For sufficiently strong coupling a phase transition appears,
driven by the unbinding of vortex pairs \cite{kos73}.
For $N>2$ as $g(\mu)$ flows toward increasingly strong coupling it
eventually leaves the regime where perturbation theory can
be considered reliable.

Above two dimensions, $d-2=\epsilon >0$, one can redo the same
type of perturbative calculation to determine the coupling constant 
renormalization.
There one finds \cite{pol75,bre76} for the Callan-Symanzik $\beta$-function  for $g$
\beq
\mu \; { \partial \, g^2 \over \partial \, \mu } \; = \; 
\beta (g) \; = \; 
\epsilon \, g^2 - { N - 2 \over 2 \pi } \, g^4 
\, + \, O \left ( g^6, \epsilon g^4 \right )    \;\; .
\label{eq:beta_nonlin}
\eeq
The latter determines the scale dependence
of $g$ for an arbitrary momentum scale $\mu$, and
from the differential equation $ \mu{ \partial g^2 \over \partial \mu } = \beta (g (\mu) )$ 
one determines how $g (\mu) $ flows as a function of
momentum scale $\mu$.
The scale dependence of $g(\mu)$ is such that if the initial
$g$ is less than the ultraviolet fixed point value $g_c$, with
\beq
g^2_c \, = \, { 2 \pi \epsilon \over N - 2 } \, + \, \dots
\label{eq:gc}
\eeq
then the coupling will flow towards the Gaussian fixed
point at $g=0$.
The new phase that appears when $\epsilon >0$ and
corresponds to a low temperature, spontaneously
broken phase with non-vanishing order parameter.
On the other hand, if $ g > g_c$ then the coupling
$g(\mu)$ flows towards increasingly strong coupling, and eventually
out of reach of perturbation theory.

The one-loop running of $g$ as a function of a sliding momentum scale $\mu=k$
and $\epsilon>0$ are obtained by integrating Eq.~(\ref{eq:beta_nonlin}), and one finds
\beq
g^2 ( \mu ) \; = \; { g^2_c \over 1 \, \pm \, a_0 \, (m^2 / \mu^2 )^{(d-2)/2} } 
\label{eq:grun_nonlin} 
\eeq
with $a_0$ a positive constant and $m$ a mass scale.
The choice of $+$ or $-$ sign is determined from whether one is
to the left (+), or to right (-) of $g_c$, in which case
$g (\mu )$ decreases or, respectively, increases as one flows away
from the ultraviolet fixed point.
The renormalization group invariant mass scale $\sim m$
arises here as an arbitrary integration constant of the renormalization group equations.
One can integrate the $\beta$-function equation 
in Eq.~(\ref{eq:beta_nonlin}) to obtain the renormalization
group invariant quantity
\beq
\xi^{-1} (g) = m(g) = {\rm const.} \; \Lambda \, 
\exp \left ( - \int^g { dg' \over \beta (g') } \right )
\label{eq:xi_beta}
\eeq 
which is identified with the correlation length appearing in
$n$-point functions.
The multiplicative constant in front of the expression on the right
hand side arises as an integration constant, and
cannot be determined from perturbation theory in $g$. 
The quantity $m=1/\xi$ is usually referred to as the
mass gap of the theory, that is the energy difference between
the ground state (vacuum) and the first excited state.

In the vicinity of the fixed point at $g_c$ one can do the
integral in Eq.~(\ref{eq:xi_beta}), using the linearized expression for the
$\beta$-function in the vicinity of the ultraviolet fixed point,
\beq
\beta (g) 
\; \mathrel{\mathop\sim_{ g \rightarrow g_c }} \;
\beta ' (g_c) \, (g - g_c) \, + \, \dots 
\label{eq:beta_lin}
\eeq
and one has for the inverse correlation length
\beq
\xi^{-1} (g) = m(g) \propto \,
\Lambda \, | \, g - g_c \, |^{\nu } \;\; ,
\label{eq:m_sigma}
\eeq
with correlation length exponent $\nu = - 1 / \beta'(g_c) \sim 1 / (d-2) + \dots$.


\vskip 20pt

\section{Nonlinear Sigma Model in the Large-$N$ Limit}
\label{sec:sigma_largen}

\vskip 10pt

A rather fortunate circumstance is provided by the fact that
in the large $N$ limit the nonlinear sigma model can be solved exactly [62-67].
This allows an independent verification of the correctness of the 
general ideas developed in the previous section, as well as a
direct comparison of explicit results for universal
quantities.
The starting point is the functional integral of Eq.~(\ref{eq:sigma_cont}),
\beq
Z = \int [ \, d \phi (x) \, ] \, 
\prod_x \, \delta \left [ \phi^2 (x) - 1 \right ] \,
\exp \left ( - \, S(\phi) \right )
\eeq
with
\beq
S(\phi) \, = \, { 1 \over 2 T } \,
\int d^d x \; \partial_\mu \phi (x) \cdot \partial_\mu \phi (x)
\eeq
and $T \equiv g^2 / \Lambda^{d-2} $, with $g$ dimensionless and
$\Lambda$ the ultraviolet cutoff.
The constraint on the $\phi$ field is implemented via an auxiliary Lagrange multiplier field $\alpha (x)$.
One has
\beq
Z = \int [ \, d \phi (x) ] \, [ d \alpha (x) ] \,
\exp \left ( - \, S(\phi,\alpha) \right )
\eeq
with
\beq
S(\phi,\alpha) \, = \, { 1 \over 2 T } \, \int d^d x \, 
\left [
( \partial_\mu \phi (x) )^2 \, + \, \alpha (x) (\phi^2 (x) -1 ) 
\right ]    \;\; .
\label{eq:sigma_cont2}
\eeq
Since the action is now quadratic in $\phi(x)$ one can integrate
over $N-1$ $\phi$-fields (denoted previously
by ${\bf \pi}$).
The resulting determinant is then re-exponentiated, and one
is left with a functional integral over the remaining
first field $\phi_1 (x) \equiv \sigma (x)$,
as well as the Lagrange multiplier field $\alpha(x)$,
\beq
Z = \int [ \, d \sigma (x) ] \, [ \, d \alpha (x) ] \,
\exp \left ( - \, S_N (\phi,\alpha) \right )
\label{eq:z_largen}
\eeq
with now
\beq
S_N (\phi,\alpha) \, = \,  { 1 \over 2 T } \, \int d^d x \, 
\left [ \,
( \partial_\mu \sigma )^2 + \alpha (\sigma^2 -1 ) \,
\right ]
\, + \, \half \, (N-1) \tr \log [ - \partial^2 + \alpha \, ] \;\; .
\label{eq:s_largen}
\eeq
In the large $N$ limit one neglects, to leading order,
fluctuations in the $\alpha$ and $\sigma$ fields.
For a constant $\alpha$ field, $ \langle \alpha (x) \rangle = m^2 $,
the last (trace) term can be written in momentum space as
\beq
\half \, (N-1) \, \int^\Lambda { d^d k \over (2 \pi)^d }  \log ( k^2 + m^2 )
\eeq
which makes the evaluation of the trace straightforward.
As should be clear from Eq.~(\ref{eq:sigma_cont2}),
the parameter $m$ can be interpreted as the mass of the
$\phi$ field.
The functional integral in Eq.~(\ref{eq:z_largen}) can be evaluated by the saddle point method,
with saddle point conditions
\bea
\sigma^2 & = & 1 - T \, (N-1) \, \Omega_d (m) 
\nonumber \\
m^2 \sigma & = & 0
\label{eq:saddle}
\eea
and the function $ \Omega_d ( m) $ given by the integral
\beq
\Omega_d ( m) \; = \; \int^\Lambda { d^d k \over (2 \pi)^d } \, { 1 \over k^2 + m^2 }  \;\; .
\eeq
The above integral can be evaluated explicitly in terms of hypergeometric functions,
\beq
\Omega_d (m) \; = \; { 1 \over 2^{d-1} \pi^{d/2} \Gamma (d/2) } 
\, { \Lambda^d  \over m^2 d } \; \,
{}_2 F_1 \left [ 1, { d \over 2 }; \, 1 + { d \over 2 } ; \, - { \Lambda^2 \over m^2 }
\right ]  \;\; .
\eeq
One only needs the large cutoff limit, $ m \ll \Lambda$, in which case one 
finds the more useful expression
\beq
\Omega_d (m) - \Omega_d (0) = m^2 \, [ \, c_1 \, m^{d-4} + c_2 \, \Lambda^{d-4} 
+ O ( m^2 \Lambda^{d-6} ) ]
\eeq
with $c_1$ and $c_2$ some $d$-dependent coefficients, given below.
From Eq.~(\ref{eq:saddle}) one notices that at weak coupling
and for $d>2$ a non-vanishing $\sigma$-field expectation
value implies that $m$, the mass of the ${\bf \pi}$ field, 
is zero.
If one sets $(N-1) \, \Omega_d (0) = 1 / T_c $, one can
then write the first expression in Eq.~(\ref{eq:saddle}) as
\beq
\sigma (T) \, = \, \pm \, [ 1 - T / T_c ]^{1/2}
\label{eq:magnetiz}
\eeq
which shows that $T_c$ is the critical coupling at
which the order parameter $\sigma$ vanishes. 
Above $T_c$ the order parameter $\sigma$ vanishes, and $m (T) $ is obtained,
from Eq.~(\ref{eq:saddle}) by solving the nonlinear gap equation
\beq
{ 1 \over T } = (N-1) 
\int^\Lambda { d^d k \over (2 \pi)^d } \, { 1 \over k^2 + m^2 }   \;\; .
\label{eq:largen_gap}
\eeq
Using the definition of the critical coupling $T_c$, one can
now write, in the interval $2<d<4$, for the common mass
of the $\sigma$ and ${\bf \pi}$ fields
\beq
m (T) \, \mathrel{\mathop\sim_{ m \, \ll \, \Lambda }} \,
\left ( { 1 \over T_c } - { 1 \over T } \right )^{1/(d-2)}  \;\; .
\label{eq:m_largen}
\eeq
This then gives for the correlation length exponent
the non-gaussian value $\nu=1/(d-2)$,
with the gaussian (Wilson-Fischer) value $\nu=1/2$ being recovered as
expected in $d=4$.
Note that in the large $N$ limit the constant of proportionality
in Eq.~(\ref{eq:m_largen}) is completely determined by the explicit
expression for $\Omega_d (m)$.
Perhaps one of the most striking aspects of the nonlinear sigma model
above two dimensions is that all particles are massless in perturbation theory,
yet they all become massive in the strong coupling phase $T>T_c$, with
masses proportional to the nonperturbative scale $m$.

Again, one can perform a renormalization group analysis,
as was done in the previous section within the context of the
$2+\epsilon$ expansion.
To this end one defines dimensionless quantities 
$g= \Lambda^{d-2} T$ and $g_c = \Lambda^{d-2} T_c$
as was done in Eq.~(\ref{eq:sigma_cont}).
Then the nonperturbative result of Eq.~(\ref{eq:m_largen}) becomes
\beq
m (g) \; \simeq \; 
c_d \cdot \Lambda \left ( { 1 \over g_c } - { 1 \over g } \right )^{1/(d-2)}
\label{eq:m_largen1}
\eeq
with the numerical coefficient given by
$c_d =[ \frac{1}{2} (d-2) \pi \vert \csc \left(\frac{d \pi }{2}\right) \vert 
]^{ { 1 \over d-2} }$.
Furthermore, one finds for the $O(N)$ $\beta$-function in the large $N$ limit
\beq
\beta (g) = (d-2) \, g^2 \, ( 1 - g^2 / g^2_c )  \;\; .
\label{eq:beta_largen}
\eeq
The latter is valid again in the vicinity of the fixed point at $g_c$,
due to the assumption, used in Eq.~(\ref{eq:m_largen}), that $m \ll \Lambda$.
Then Eq.~(\ref{eq:beta_largen}) gives the momentum dependence of the
coupling at fixed cutoff, and upon integration one finds
\beq
{ g^2 (\mu) \over g^2_c } = { 1 \over 1 - c \,( \mu_0 / \mu )^{d-2} } \approx 
1 + c \, ( \mu_0 / \mu )^{d-2} + \dots
\eeq
with $c \, \mu_0^{d-2}$ a dimensionful integration constant.
The sign of $c$ then depends on whether one is on the right ($c>0$) or
on the left ($c<0$) of the ultraviolet fixed point at $g_c$.
At the fixed point $g_c$ the $\beta$-function
vanishes and the theory becomes scale invariant.
Moreover, one can check again that $\nu = - 1/\beta'(g_c)$
where $\nu$ is the exponent in Eq.~(\ref{eq:m_largen}).


\vskip 20pt

\section{Hartree-Fock Method for the Nonlinear Sigma Model}
\label{sec:sigma_hf}

\vskip 10pt

\begin{figure} 
     \begin{center}
       \includegraphics[width=0.75\textwidth]{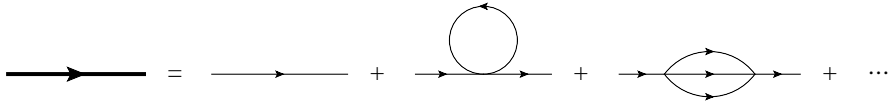}
     \end{center}
     \caption{Perturbative expansion for the dressed propagator of a self-interacting scalar field.}
     \label{fig:scalar_loop}
\end{figure}

The large $N$ saddle point equation of Eq.~(\ref{eq:largen_gap}) 
is in fact identical to the Hartree-Fock approximation
for the self-energy, where the bare propagator in the loop is replaced by
the dressed one, with a mass parameter $m$ to be determined by the 
Hartree-Fock self-consistency conditions.
Figure \ref{fig:scalar_loop} shows the typical perturbative expansion for the proper
self-energy of a self-interacting scalar field.


\begin{figure}
	\begin{center}
       \includegraphics[width=0.75\textwidth]{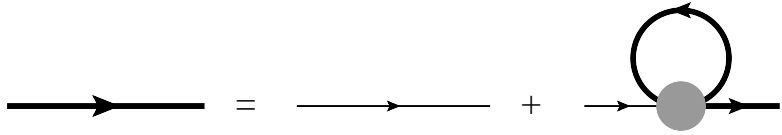}
	\end{center}
	\caption{Dyson's equation for the scalar field propagator.}	
	\label{fig:scalar_dyson}
\end{figure}

\begin{figure}
	\begin{center}
		\includegraphics[width=0.40\textwidth]{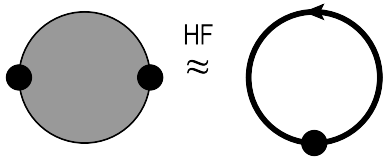}
	\end{center}
	\caption{Self-consistent Hartree-Fock approximation for the self-energy of the scalar field.}
	\label{fig:scalar_hf}
\end{figure}

To derive the Hartree-Fock approximation, one can first write down Dyson's equations
for the dressed propagator and the dressed vertices, as shown in 
Figure \ref{fig:scalar_dyson}.
The Hartree-Fock approximation to the self-energy is then obtained by replacing the bare propagator with a dressed one in the lowest order loop diagram, as shown in Figure \ref{fig:scalar_hf}.
There the dressed scalar propagator is to be determined self-consistently from the solution 
of the nonlinear Hartree-Fock equations.
It should be noted that in the lowest order approximation the loop integrals
involve the dressed
scalar propagator $ \propto (k^2+m^2)^{-1}$,  while the scalar vertex is still the bare one. 
This is a characteristic feature of the Hartree-Fock approximation both in many-body
theories \cite{fet71} and in QCD \cite{alk01,rob94}.


One important aspect that needs to be brought up at this stage is the dependence
of the results on the specific choice of ultraviolet cutoff.
In the foregoing discussion the momentum integrals were cut off, in magnitude, at some
large virtual momentum $\Lambda$, which resulted in an explicit expression for the gap equation 
which reflected, at low momenta, the original spherical symmetry of the underlying cutoff.
The latter is nevertheless not the only possibility.
An alternative procedure would be the introduction of an underlying hypercubic (or other) 
lattice, with lattice spacing $a$ and for which then the momentum cutoff is $\Lambda = \pi / a $. 
One more possibility would be the use of dimensional regularization, where ultraviolet
divergences re-appear as simple poles in the dimension $d$.
The detailed choice of ultraviolet cutoff would then affect the form of the $\Lambda$ terms 
in the gap equation.
It is nevertheless well known, and expected on the basis of very general and well-established renormalization group arguments (based mainly on the nature of what are referred to as 
irrelevant operators) that the main effect would go into a shift of the non-universal 
critical coupling $T_c$, leaving the universal exponents and their dependence
on the dimension $d$ and on the $N$ of $O(N)$ unchanged.
See for example [1-6], the comprehensive monographs [7-11], and the many more references therein.

In view of the later discussion on quantum gravity,
it will be useful here to next look explicitly at a few individual cases, mainly as far as the 
dimension $d$ is concerned.
These follow from the general expression for the loop integral given above
\beq
S_d \; \int_0^\Lambda dk \, k^{d-1} \, { 1 \over k^2 + m^2 }
\label{eq:int_sigma}
\eeq
with $ S_d = 1 / 2^{d-1} \, \pi^{d/2}  \, \Gamma ( d/2 ) $.
One finds for the integral itself in general dimensions
\beq
I (\Lambda,m , d ) \; \equiv \; 
\int_0^\Lambda \, dk \, k^{d-1} \, { 1 \over k^2 + m^2 } \; = \; 
{\Lambda^d \over d \, m^2 } \,\,  {}_2 F_1 ( 1,{d \over 2}, { 2 + d \over 2}, - { \Lambda^2 \over m^2 } )
\label{eq:integral_sigma}
\eeq
with ${}_2 F_1 ( a,b,c,z) $ the generalized hypergeometric function.
The general nonlinear gap equation for $m$ in $d$ dimensions then reads
\beq
(N-1) \, T \cdot S_d \cdot {\Lambda^d \over d \, m^2 } \,\,
{}_2 F_1 ( 1,{d \over 2}, { 2 + d \over 2}, - { \Lambda^2 \over m^2 } ) \, - \, 1 \; = \; 0  \;\; .
\label{eq:gap_sigma}
\eeq
Here one recognizes that the only remnant of the large $N$ expansion is the overall
coefficient $\beta_0 = N-1$; perhaps a better, physically motivated, choice would have been
a factor of $N-2$, since $N=2$ is known to be the special case for which
the mass gap disappears in the weak coupling, spin wave phase in $d=2$.
\footnote{
The self-consistent Hartree-Fock gap equation for $m$ in the nonlinear sigma model 
shares some analogies with the gap equation for a superconductor.
There one finds, in the Hartree-Fock approximation, the following nonlinear
gap equation \cite{agd62,fet71},
\beq 
g \, N(0) \, \int_0^{ \hbar \omega_D } 
{ d \xi \over ( \xi^2 + \Delta^2 )^{\half}  }
\, \tanh { ( \xi^2 + \Delta^2 )^{\half}  \over 2 k_B \, T }
\, = \, 1
\label{eq:bcs_gap}
\eeq
where $g$ is the electron-phonon coupling constant,
$\omega_D$ the upper phonon Debye (cutoff) frequency,
$\Delta$ is the electron energy gap at the Fermi surface, $T$ the temperature
and $\xi$ a shifted energy variable defined as
\beq
\xi \, = \, { \hbar^2 \, k ^2 \over 2 \, m  } \, - \, \mu  \;\; .
\eeq
In the above expression $\mu $ is the chemical potential, and the quantity 
$N(0) = m \, k_F / 2 \pi^2 \hbar^3 $ indicates the density of states for 
one spin projection near the Fermi surface, with $k_F$ the Fermi wavevector.
The above equation then leads, in close analogy to what is done
in the nonlinear sigma model, to an estimate for the critical temperature
at which the electron gap $\Delta$ vanishes
\beq
k_B \, T_c \, = \, { 2 e^{\gamma} \over \pi } \, \hbar \, \omega_D \,
e^{ - 1 / N(0) g } \; \; .
\eeq
Note that the role of the ultraviolet cutoff here is played by $\omega_D$,
and that the critical temperature is non-analytic in the electron-phonon 
coupling $g$.
Furthermore the role of the mass gap $m$ in the nonlinear sigma model is played
here by the energy gap at the Fermi surface $\Delta$.
The condensed spin zero electron Cooper pairs then act as massless Goldstone bosons,
and later induce a dynamical Higgs mechanism in the presence of an external
electromagnetic field, commonly referred to as the Meissner effect.
}


It now pays to look at some specific dimensions individually, just
as will be done later in the case of quantum gravity.
Then specifically, for dimension $d=2$, one has 
\beq
\int_0^\Lambda \, dk \, k \, { 1 \over k^2 + m^2 } \; = \; 
\half \, \log \left [ { m^2 + \Lambda^2 \over m^2 } \right ]
\eeq
leading to the $d=2$ gap equation 
\beq
(N-1) \; {T \over 4 \pi }  \,  \log \left ( { m^2 + \Lambda^2 \over m^2 } \right ) \; = \; 1 \;\; .
\eeq
The solution is given by
\beq
m \; = \; \Lambda \, \left [ e^{ 4 \pi \over (N-1) \, T } -1 \right ]^{- \half } 
\; = \; 
\Lambda \, e^{ - { 2 \pi \over (N-1) \, T } } \left ( 1 + \half \, e^{ - { 4 \pi \over (N-1) \, T } } \, + \, \cdots \right )
\eeq
so that here the constant of proportionality between the mass gap $m$ and
the scaling parameter $ \Lambda \, e^{ - { 2 \pi \over (N-1) T } } $ is exactly one for weak
coupling $T = g^2 \ll 1 $.
Then the above result corresponds to a correlation 
length exponent $\nu = \infty $ in $d=2$.
On the other hand, in the opposite strong coupling limit $ T = g^2 \gg 1 $ one has
\beq
m \; \simeq \; { g \, \sqrt{ N-1}  \over 2 \, \sqrt{\pi}  } \; \Lambda 
\eeq
so that, as expected, the correlation length $\xi = 1 / m $ approaches zero in this limit.


Moving up one dimension, in three dimensions, $d=3$, one has for the basic integral
\beq
\int_0^\Lambda \, dk \, k^2 \, { 1 \over k^2 + m^2 } \; = \; 
\Lambda - m \arctan \left ( { \Lambda \over m } \right ) 
\eeq
leading to the $d=3$ gap equation 
\beq
(N-1) \; {T \over 2 \pi^2 }  \,  
\left [ \Lambda - m \arctan \left ( { \Lambda \over m } \right )  \right ]
\; = \; 1
\eeq
with critical point at
\beq
T_c \; = \; { 2 \pi^2 \over (N-1)  \, \Lambda }
\eeq
for which the mass gap $m=0$.
In the vicinity of the critical point one can solve explicitly for the mass gap $m$
\beq
m \; = \; { 4 \pi \over (N-1) \, T_c^2 } \, \vert \, T-T_c \, \vert \, + \, 
{ 4 \, ( \pi^2 -4 ) \over (N-1) \, \pi \, T_c^3 } \, \vert \, T-T_c \, \vert^2  \, + \, \cdots
\eeq
so that in $d=3$ the universal correlation length exponent is $\nu=1$.
Equivalently, in terms of the dimensionless coupling $g$, for which $ T = g^2 / \Lambda $,
one has $g_c = \sqrt{2} \, \pi / \sqrt{ N - 1 } $ and therefore
\beq
m \; = \; \Lambda \; { 8 \pi \over (N-1) \, g_c^3 } \; \vert  g - g_c \vert \, + \, \cdots \; .
\eeq


Again moving up in dimensions, one finds that four dimensions ($d=4$) represents a marginal case 
and one obtains
\beq
\int_0^\Lambda \, dk \, k^3 \, { 1 \over k^2 + m^2 } \; = \; 
\half \left [ \Lambda^2 \, + \, m^2 \, \log \left ( { m^2 \over m^2 + \Lambda^2 } \right ) \right ]
\eeq
leading to the $d=4$ gap equation 
\beq
(N-1) \, {T \over 16 \pi^2 }  \,  
\left [ \Lambda^2 \, + \, m^2 \, \log \left ( { m^2 \over m^2 + \Lambda^2 } \right ) \right ]
\; = \; 1
\eeq
with critical point
\beq
T_c \; = \; { 16 \pi^2 \over (N-1)  \, \Lambda^2 } \;\; ,
\eeq
and therefore $g_c = 4 \pi / \sqrt{N-1}  $ where $m=0$.
The four-dimensional case is slightly more complex, and here one has to
determine the solution to the gap equation recursively.
In the vicinity of the critical point one finds for the mass gap $m$
\beq
m \; = \; \Lambda \, { 4 \, \sqrt{2} \pi \over (N-1)^{1/2} \, g_c^{3/2}  } \; 
{  \vert g - g_c \vert^{1/2} \over \left ( - \log \vert g - g_c \vert \right )^{1/2}  }
 \, + \,  \cdots \;\; ,
\label{eq:mass_nonlin_4d}
\eeq
so that in $d=4$ the universal correlation length exponent is given by the Landau theory
value $\nu=1/2$, up to logarithmic corrections.
This last result is in agreement with triviality arguments
for scalar field theories in and above four dimensions \cite{wil72,wil72a,wil73}.
Also, the logarithmic correction has the right form with the correct power $-1/2$, in 
agreement with the exact universal result for the Ising model (given here by the $N$=1 case) 
in four dimensions \cite{pfe76}.


Going up one more dimension, in five dimensions ($d=5$) one has for the integral
\beq
\int_0^\Lambda \, dk \, k^4 \, { 1 \over k^2 + m^2 } \; = \; 
\third \, 
\left [ \, \Lambda^3 \, - 3  \, m^2 \, \Lambda + 3 \, m^3 \, \arctan \left  ( { \Lambda \over m } \right ) \right ] \eeq
leading to the $d=5$ gap equation 
\beq
(N-1) \, {T \over 36 \pi^3 }  \,  
\left [ \, \Lambda^3 \, - 3  \, m^2 \, \Lambda + 3 m^3 \, \arctan \left  ( { \Lambda \over m } \right ) 
\right ] \; = \; 1
\eeq
with critical point at
\beq
T_c \; = \; { 36 \, \pi^3 \over (N-1)  \, \Lambda^3 } \;\; ,
\eeq
and therefore $g_c = 6 \, \pi^{3/2} / \sqrt{N-1}  $, at which point again $m=0$.
In the vicinity of the critical point one finds for the mass gap $m$
\beq
m \; = \; \Lambda \, { 2 \, 6^{1/6} \, \pi \over (N-1)^{2/3} \, g_c^{7/6}  } 
\; \vert g - g_c \vert^{1/2}  \, + \,  \cdots
\eeq
so that in $d=5$ (and above) the universal correlation length exponent stays at
$\nu=1/2$, again in agreement with the Landau theory prediction for all scalar 
field theories above $d=4$ \cite{wil72,wil72a,wil73}.


More generally, the last set of results agree with the general formula for 
the critical point $T_c$, valid for both $d<4$ and $d \geq 4$,
\beq
T_c \; = \; { (d-2) \, 2^{d-1} \, \pi^{d \over 2} \, 
\Gamma \left ( { d \over 2 } \right ) \over N-1 } \; \Lambda^{2-d}
\eeq
and similarly for the dimensionless critical coupling $g_c = \sqrt{T_c} /  \Lambda^{(2-d)/2} $.
A dimensionless critical amplitude $A_m$ can be defined by 
\beq
m \; = \;\; \mathrel{\mathop\sim_{g \rightarrow g_c }} \;\;
A_m \, \Lambda \, \vert g - g_c \vert^{1/(d-2)} \, + \, \cdots
\eeq
and is given for $d<4$ explicitly by the expression
\beq
A_m \; = \; \left ( 
{ 2^{d+1} \, \pi^{{ d \over 2}-1} \Gamma ( { d \over 2 } ) \, 
\sin \left ( { \pi d \over 2 } \right ) \over g_c^3 \, (N-1) } \right )^{1 \over d-2 } \;\; .
\eeq
For $d>4$ the amplitude can be computed explicitly as well, and is given instead by
\beq
A_m \,=\, 2^{ 3d-4 \over 2 (d-2)} \, (d-2)^{ 4-d \over 2(d-2)} \, (d-4)^{\half} \, g_c^{ d+2 \over 2(2-d) } \, n^{- { 1 \over d-2 }} \pi^{d \over 2(d-2)} \, \Gamma ( { d \over 2 } )^{ 1 \over d-2 }
\eeq
which shows again that $d=4$ is indeed a special case, and needs to be treated with some care.
One also notes that both the $d<4$ and the $d>4$ amplitudes vanish as one approaches $d=4$ due
to the infrared divergence in this case, which leads to the $log$ correction described above.

The previous, rather detailed, discussion shows that the Hartee-Fock approximation (and, in this case, the equivalent large-$N$ limit), based on a single one loop tadpole diagram, leads to a number of basically correct and nontrivial analytical results in various dimensions, which we proceed to 
enumerate here.
The first one is the fact that it correctly predicts the absence of a phase transition, and thus asymptotic freedom in the coupling $g$, for any $N>1$ in two dimensions. 
This result is indeed known to be correct up to $N=2$, the latter representing the special case of the Kosterlitz-Thouless vortex unbinding transition in the two dimensional planar model.
The beta function in this case becomes then exactly correct for weak coupling, if one replaces the saddle point value $N-1$ by $N-2$ in the relevant expressions.

Secondly, in three dimensions the theory is known not to be perturbatively renormalizable.
Nevertheless the Hartree-Fock approximation correctly predicts the existence of a critical point (a phase transition) at a finite $T_c$, and the correspondingly modified scaling relations.
In renormalization group language a critical point corresponds to a nontrivial fixed point
associated with the renormalization group trajectories, here in the dimensionless relevant coupling $g$.
The critical exponent $\nu=1$ is somewhat higher than the accepted value for the Heisenberg model $N=3$ ($\nu=0.710 $)  \cite{gui98}, nevertheless not too far off, given the crudeness of the approximation, based on a single loop diagram.

Furthermore, for dimension $d$ between two (lower critical) and four (upper critical), one
has 
\beq
\nu = { 1 \over d - 2} \;\;\;\;\;\; 
\eta = 0  \;\;\;\;\;\; 
\gamma = { 2 \over d - 2}
\eeq
where $\eta$ is the field (or propagator) anomalous dimension, $G(p) \sim 1 / p^{2 - \eta} $ 
in the vicinity of the nontrivial fixed point,
and $\gamma$ the zero-field magnetic susceptibility exponent, $\chi \sim \xi^{\gamma / \nu }$,
again in the vicinity of the nontrivial fixed point at $g_c$.
In four dimensions the Hartree-Fock approximation correctly reproduces the expectation of
mean field exponents ($\nu=1/2$ here) up to a logarithmic correction. 
This is not surprising, since some sort of mean field theory is incorporated into
the very nature of the Hartree-Fock approximation.
Above four dimensions the results again reflect, correctly, the expectation that mean field theory exponents, and in particular $\nu=1/2$, should apply to all $d \ge 4 $, and give an explicit analytic expression for the mass gap $m$, the critical coupling $T_c$ and the corresponding
amplitude $A_m$ as a function of $d$.
Thus the Hartree-Fock approximation as discussed here gives correctly for the upper
critical dimension of the $O(N)$ nonlinear sigma model $d=4$ independent of $N$, 
whereas the lower critical dimension here is correctly $d=2$, at least for $N>2$.

To conclude this section, a few words should be spent on the physical interpretation of the
mass gap parameter $m$.
It is possible to describe its properties either within the context of Lorentzian quantum field theory
with its generating functional and vacuum expectation values, or alternatively in 
the context of statistical
mechanics and Euclidean quantum field theory, with its correlation functions and
thermodynamic averages. 
The two descriptions are of course expected to be equivalent and complemetary, 
related to each other by a Wick rotation.
So, when one refers to the mass gap, it means the gap in energy between the first
excited stated and the ground state, $m = E_1 - E_0 $ as derived from the quantum
Hamiltonian or transfer matrix describing, in this case, the nonlinear sigma model. 
Its relationship with the Euclidean correlation length $\xi$ is provided by the Lehman
representation (i.e. completeness) for the two-point function, for example, 
which then gives $m=1/ \xi$.
In addition, $\xi$ and thus $m$ are related by scaling to a multitude of other
observables, such as the order parameter or magnetization $\langle \sigma \rangle$ 
in the low temperature phase.
As an example, in the nonlinear sigma model 
$\langle \sigma \rangle \sim 1 / \xi^{\beta / \nu }$ by renormalization group scaling,
where $\beta$ is the magnetic exponent ($\approx 1/3$ in $d=3$) and $\nu$ the 
correlation length exponent.
This shows more generally that the mass gap is in some way directly related to the vacuum 
expectation value of the field, at least in the ordered phase.
For Gaussian fields $\langle \sigma \rangle \; \sim 1/ \xi $ in four dimensions in the 
ordered phase, simply by dimensional arguments.


\section{Gauge Theories} 
\label{sec:gauge}   

\vskip 10pt

\begin{figure}
	\begin{center}
		\includegraphics[width=0.85\textwidth]{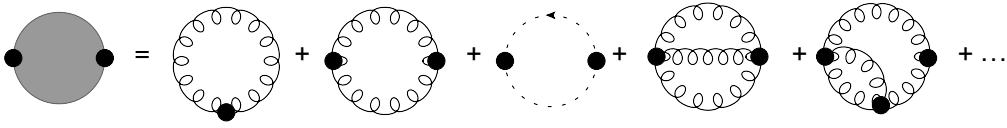}
	\end{center}
	\caption{Pure Yang-Mills vacuum polarization contributions to lowest order in 
                perturbation theory. The dashed line represents the ghost contribution.}
	\label{fig:ym_pert}
\end{figure}

\begin{figure}
	\begin{center}
		\includegraphics[width=0.75\textwidth]{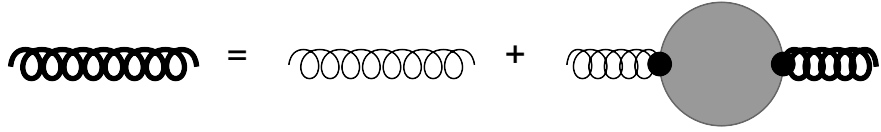}
	\end{center}
	\caption{Proper vacuum polarization in Yang-Mills theories (the large gray shaded blob).}
	\label{fig:ym_blob}
\end{figure}

Before tackling the most complex case of quantum gravity, it will be useful to first explore 
the implications of these ideas to a new and slightly more complex, but still familiar 
case - the $SU(N)$ gauge theory.
In the case of $SU(N)$ Yang-Mills theories one can follow a similar line of analysis to
develop the Hartree-Fock approximation.
Overall, the strategy is again to rely on well-known results for one loop diagrams, suitably
modified by the insertion of an effective, dynamically generated mass parameter.
The latter is then determined self-consistently by a suitable nonlinear gap equation.
There are a number of aspects which are quite similar to the nonlinear sigma model
case, which will be helpful here in vastly streamlining the discussion.
In particular, in gauge theories the gluon propagator is modified by the gauge and matter 
vacuum polarization contribution $\delta^{ab} \Pi_{\mu\nu} (p) $ with
\beq
\Pi_{\mu\nu} (p) \; = \; ( p^2 \, \eta_{\mu\nu} - p_\mu \, p_\nu ) \, \Pi (p^2)
\eeq
with $p_\mu \Pi^{\mu\nu} (p) =0 $.
In the Fermi-Feynman gauge  the gluon propagator then becomes
\beq
D_{\mu\nu}^{ab} (p) \; = \; { \delta^{ab} \, \eta_{\mu\nu} \over p^2 \, [ 1 + \Pi (p^2) ] } \;\; .
\eeq
In perturbation theory, and to all orders, the gluon stays massless as a consequence
of gauge invariance, with
$ p^2 \, \Pi (p^2) \; \mathrel{\mathop\sim_{ p \rightarrow 0 }} \,  0 $
(here we will assume no spontaneous symmetry breaking via the Higgs mechanism).
In more detail, the relevant one loop integral has the form
\beq
\int { d^d k \over ( 2 \pi )^d }  \,
{ V (p,-k-p,k ) \cdot V (-p,-k,k+p) \over k^2 \, (k+p)^2 }
\eeq
where the dot here indicates a dot product over the relevant Lorentz indices
associated with the three-gluon, matter or ghost vertices.
Additional factors involve a symmetry factor $\half$ for the diagram, 
the gauge coupling $g^2$ weight, and an overall group theory color factor.
Figure \ref{fig:ym_pert} shows the lowest order Feynman diagrams contribution to the vacuum polarization tensor in gauge theories, Figure \ref{fig:ym_blob} the equation relating the bare gluon
propagator to the dressed one via the proper vacuum polarization contribution.


\begin{figure}
	\begin{center}
		\includegraphics[width=0.75\textwidth]{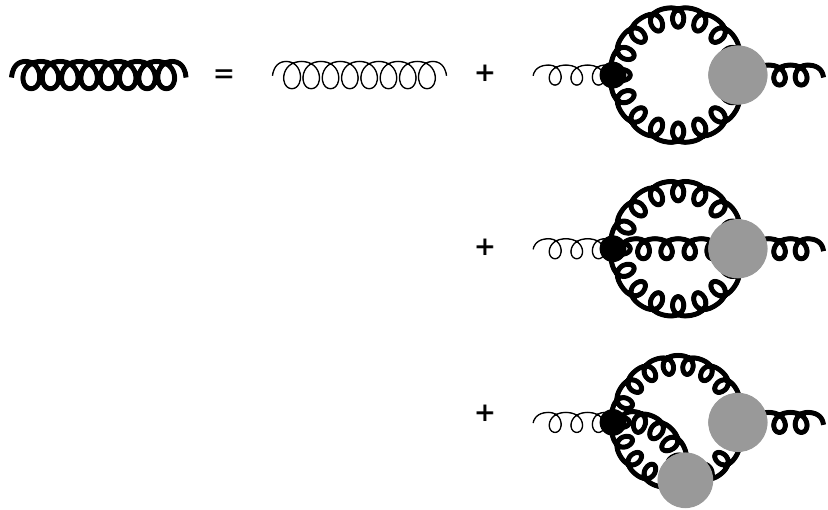}
	\end{center}
	\caption{Dyson's equations for the dressed Yang-Mills boson propagator. 
               Ghost  contributions have been omitted from the diagrams.}
	\label{fig:ym_dyson}
\end{figure}

\begin{figure}
	\begin{center}
		\includegraphics[width=0.75\textwidth]{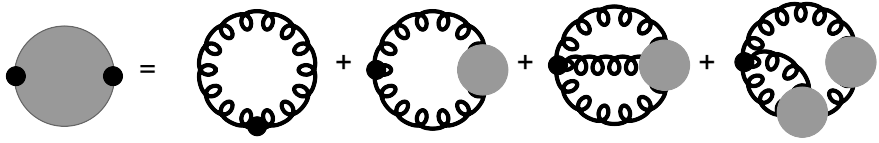}
	\end{center}
	\caption{Dyson's equations for the Yang-Mills gauge boson vacuum polarization tensor.}
	                \label{fig:ym_dyson_self}
\end{figure}

Figure \ref{fig:ym_dyson} then represents Dyson's equation for the dressed gluon propagator (as written out in terms of the bare propagator and bare and dressed vertices, while
Figure \ref{fig:ym_dyson_self} illustrates Dyson's equation for the proper gluon vacuum polarization term.


\begin{figure}
	\begin{center}
		\includegraphics[width=0.40\textwidth]{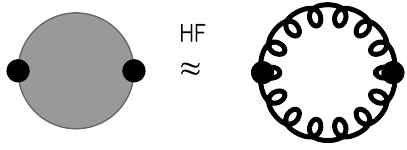}
	\end{center}
	\caption{Self-consistent Hartree-Fock approximation for the gauge boson proper self-energy.}
	\label{fig:ym_hf}
\end{figure}

Finally, Figure \ref{fig:ym_hf} illustrates the Hartree-Fock equation for gauge theories.
Consequently, in the Hartree-Fock approximation (and in close analogy to the treatment
of the nonlinear sigma model described previously, specifically in Section \ref{sec:sigma_hf}) 
one is left with the evaluation of the following one loop integral,
\beq
S_d \; \int_0^\Lambda dk \, k^{d-1} \, { k^{ 2 \sigma}  \over (k^2 + m^2)^2 }  \;\; ,
\eeq
with again $ S_d = 1 / 2^{d-1} \, \pi^{d/2}  \, \Gamma ( d/2 ) $ and $\sigma = 0$ in the gauge
theory case.
Here $\Lambda$ is an ultraviolet cutoff, such as the one implemented
in Wilson's lattice gauge theory formulation, for which $\Lambda \sim \pi / a $ with
$a$ the lattice spacing.
Note the insertion of a mass parameter $m$, to be determined later self-consistently by
the gap equation given below.


One finds for the integral itself in general dimensions the explicit result
\beq
\int_0^\Lambda \, dk \, k^{d-1} \, \, { 1 \over (k^2 + m^2)^2 }
 \; = \; 
{\Lambda^d  \over 2 \, d  \, m^2 \, (m^2 + \Lambda^2 ) } \, + \, 
\left ( {1 \over d} - { 1 \over 2 } \right ) \, 
{\Lambda^d \over d \, m^4 } \,\,  
{}_2 F_1 ( 1, {d \over 2}, 1 + {d \over 2}, - { \Lambda^2 \over m^2 } )
\label{eq:integral_ym}
\eeq
with ${}_2 F_1 ( a,b,c,z) $ the generalized hypergeometric function.
The general gap equation for the dynamically generated mass scale $m$ then reads
\beq
g^2 \; \beta_0 \; S_d \, \left [
{\Lambda^d \over 2 \, d \, m^2 \, (m^2 + \Lambda^2 ) } \, + \, 
\left ( {1 \over d} - { 1 \over 2 } \right ) \, 
{\Lambda^d \over d \, m^4 } \,\,  
{}_2 F_1 ( 1, {d \over 2}, { 1 + d \over 2}, - { \Lambda^2 \over m^2 } ) 
\right ] \; = \; 1  \;\; ,
\label{eq:gap_eq_ym_d}
\eeq
with $g$ the gauge coupling and $\beta_0$ an overall $N$-dependent numerical
coefficient resulting from combined group theory weights, Lorentz traces and diagram
symmetry factors.


As in the case of the nonlinear sigma model discussed previously, it will
be instructive to look in detail at individual dimensions.
In four dimensions ($d=4$) the gap equation reduces to
\beq
{ 1 \over 2 } \, g^2 \; \beta_0 \; S_4 \, \left [ \,
\log \left (  1+ { \Lambda^2 \over m^2 } \right ) \, - \, 
{ \Lambda^2 \over m^2 + \Lambda^2 }  \, \right ]
\; = \; 1   \;\; .
\label{eq:gap_eq_ym_4}
\eeq
The solution to the above equation is
\beq
m \; = \; \Lambda \, e^{- \half } \; e^{- { 1 \over 2 \, \beta_0 \, g^2 } } 
\; = \; \Lambda \, e^{- \half} \; e^{- { 24 \, \pi^2  \over 11 \, N \, g^2 } }  \;\; .
\label{eq:m_lambda}
\eeq
Here at the end the correct expression $\beta_0  = 11 N / 3 $ for $SU(N)$ gauge theories  
was inserted, as obtained from one loop perturbation theory.
The above explicit results then implies, within the current Hartree-Fock approximation,
that the mass gap $m$ is by a factor $ e^{-\half} = 0.6065 $ smaller
than the scaling violation parameter $ \Lambda \, \exp ( - 1 / 2 \, \beta_0 \, g^2 ) $,
with the latter appearing on the r.h.s. of Eq.~(\ref{eq:m_lambda}).
Note that one could have obtained the overall coefficient of the one
loop contribution, as used here in the Hartree-Fock approximation, 
from the Nielsen-Hughes formula \cite{nie81,hug80}.
There the one-loop $\beta$-function contribution coefficient 
arising from a particle of spin $s$ running around the loop is given generally by 
$ \beta \, = \, - (-1)^{2 s} \, [ (2 s)^2 - \third ] $,
with the two individual competing contributions arising from a generalization of
Pauli diamagnetism and Landau paramagnetism in four spacetime dimensions.
For spin $s=1$ the above formula then gives indeed for the pure $SU(N)$ gauge theory in four
dimensions the expected result, $\beta_0  = 11 N / 3 $ (with, in this case, an 
overall negative sign for the beta function, correctly taken into account previously).

The renormalization group running of the gauge coupling $g(\mu)$ again follows from 
the requirement that the mass gap $m$ be scale independent,
\beq
\mu \, { d \, m \over d \, \mu } \; = \;  0 \; = \;
\mu \, { d \, \over d \, \mu } \, \left ( 
\mu \, e^{- \half} \, e^{- { 1 \over 2 \, \beta_0 \, g ( \mu ) ^2 } }  
\right )  \;\;  ,
\eeq
which then gives the expected asymptotic freedom running of $g(\mu)$ in the vicinity of 
the ultraviolet fixed point at $g=0$,
\beq
g^2 ( \mu ) \; = \;  { 1 \over  2 \, \beta_0 \, [ \, \log ( \mu /  m )  - 1 \, ] }    \;\; .
\label{eq:grun_ym_4d}
\eeq
One observes that the quantity $m$ here plays a role analogous to the 
$\Lambda_{\bar MS} \approx 330 \text{MeV} $ 
parameter used, for example, to describe deep inelastic scattering in QCD.
For short enough distances the result of Eq.~(\ref{eq:grun_ym_4d}) implies that the static
potential between quarks behaves as $ V ( r ) \sim 1 / r \, log ( r ) $, a Coulomb potential with a weak
quantum logarithmic correction, of course a well-known result already from perturbation theory.
It should be noted also that so far the gauge theory result in $d=4$ looks rather similar to the 
$d=2$ result for the  nonlinear sigma model (for $N>2$), in the sense that both models 
have a mass gap that
vanishes only at $g=0$, and also exhibit asymptotic freedom in the relevant coupling.
This fact is of course already well known from the simple, direct application of perturbation theory;
nevertheless it is encouraging that both of those features are correctly reproduced by the lowest
order Hartree-Fock approximation.


Gauge theories are mostly relevant in four space-time dimensions. 
As a purely academic exercise one can nevertheless look at $d=5$, 
where Yang-Mill theories - just like the nonlinear sigma model for $ d>2 $ - are
not perturbatively renormalizable.
In this case one needs the integral 
\beq
\int_0^\Lambda \, dk \, k^5 \; { 1 \over (k^2 + m^2)^2 }
 \; = \; 
\Lambda \, + \, {m^2 \, \Lambda \over 2 \, ( m^2 + \Lambda^2 ) } \, - \, 
{3 \over 2} \, m \, \arctan \left ( { \Lambda \over m } \right )  \;\; .
\eeq
For small mass gap parameter $m$ the gap equation in five dimensions then reads
\beq
g^2 \; \beta_0 \; S_5 \, \left [ \, \Lambda \, - \, { 3 \pi \over 4 } \, m \, \right ]
\; = \; 1 \;\; .
\label{eq:gap_eq_ym_5}
\eeq
The solution for $ g > g_c $  is then given by
\beq
m \; = \; \Lambda \, \cdot { 8 \over  3 \pi g_c } \, ( g - g_c ) 
\, + \, {\cal O} \left (  (g-g_c)^2 \right ) \;\; .
\eeq
There is a nontrivial fixed point at $ g_c^2 = 12 \pi^3 / \Lambda $, and the correlation length exponent 
is $\nu=1$, just like for the $d=3$ nonlinear sigma model discussed previously.
For the gauge theory case one expects that the critical point at $g_c$ separates a weak coupling
Coulomb phase from a confining strong coupling case.
Then, as $d$ approaches four from above, the weak coupling massless gluon Coulomb phase 
disappears.

The renormalization group running of the gauge coupling $g(\mu)$ close to the nontrivial 
ultraviolet fixed point then follows again from the requirement that the mass gap $m$ be scale 
independent, $ \mu \, { d \, m \over d \, \mu } \; = \;  0 $.
In the vicinity of the five-dimensional ultraviolet
fixed point, and within the strong coupling phase ($ g > g_c $) one obtains
\beq
g ( \mu ) \; = \; g_c \, 
\left [ \, 1 \, + \, { 3 \pi \over 8 } \cdot \left ( { m \over \mu } \right ) 
\, + \, {\cal O} \left (  \left ( { m \over \mu } \right )^2 \right  )  \right ] \;\; .
\label{eq:grun_ym_5d}
\eeq
Note that, in this case, the power associated with the running in momentum space 
is determined again by the exponent $1 / \nu = 1 $, and
that the coefficient of the running term here is in fact independent of $\beta_0$.
The result of Eq.~(\ref{eq:grun_ym_5d}) in turn implies that in real space 
the effective coupling constant initially grows linearly with distance in the strong coupling phase, 
$ g ( r ) \sim r / \xi $.
At this stage one would conclude that the gauge theory result in $d=5$ bears some similarity
to the $d=3$ result for the nonlinear sigma model, in the sense that both models have a critical point
located at some finite value of the bare coupling, and have an exponent $\nu=1$, at least within
the Hartee-Fock approximation.


In fact, it is possible to get a general yet simple analytical result from the Hartree-Fock
result for larger $d$ as well, based on a general expression for the momentum integral.
For a large enough dimension, $d>6$, one has
\beq
g_c^2  \; = \; { (d-4) \, 2^{d-1} \, \pi^{d/2} \, \Gamma (d/2) \over \beta_0 \, \Lambda^{d-4} } \;\; .
\eeq
The solution for $m$ is then, for $ g > g_c $ and $d>6$, 
\beq
m \; = \; \Lambda \, \left ( { d-6 \over (d-4) \, g_c } \right )^{1/2} \, ( g - g_c )^{1/2}  
\, + \, {\cal O} \left (  (g-g_c)^{3/2} \right ) \;\;  ,
\eeq
which implies $\nu=1/2$ for any $d>6$, just like the case $d>4$ for the nonlinear 
sigma model discussed earlier.
Note that six dimensions is the borderline case here, with logarithmic corrections to a pure power
law behavior, in analogy to the $d=4$ case for the nonlinear sigma
model, as given in Eq.~(\ref{eq:mass_nonlin_4d}).

Again, the renormalization group running of $g(\mu)$ follows
from the requirement that the mass gap $m$ be scale independent, and
in the vicinity of the ultraviolet fixed point, and in the strong coupling phase $ g > g_c $, 
one finds here
\beq
g ( \mu ) \; = \; g_c \, 
\left [ 1 \, + \, { d-4 \over d-6 } \, \left ( { m \over \mu } \right )^2  
\, + \, {\cal O} \left (  \left ( { m \over \mu } \right )^4 \right  )  \right ] \;\; .
\label{eq:grun_ym_d}
\eeq
It follows that the power associated with the running in momentum space 
is always given by the exponent $1 / \nu = 2 $ for $d>6$.
The result of Eq.~(\ref{eq:grun_ym_d}) in turn implies that in real space 
the effective coupling constant initially grows with the square of the distance, above $d=6$ and
in the strong coupling phase, $ g ( r ) \sim ( r / \xi )^2 $.
Note that, when written in this form, the coefficient of the running
term is in fact independent of the magnitude of $\beta_0$.

Overall, the general conclusion here is that the gauge theory result in $d>6$ looks rather similar to the 
$d>4$ result for the nonlinear sigma model, in the sense that both models have a critical point
at some value of the bare coupling, and have exponent $\nu=1/2$, at least within
the Hartee-Fock approximation used here.
In other words, the Hartree-Fock approximation gives for the upper
critical dimension of gauge theories $d=6$ independent of $N$, whereas the lower critical
dimension here is $d=4$, again independent of $N>1$.
As a comparison, we note that some time ago it was suggested that the upper critical 
dimension for gauge theories is either $d=8$ based on the theory of random surfaces \cite{par79}, 
or possibly $d=\infty$ \cite{dro79} when considering the strong coupling expansion of lattice 
gauge theories at large $d$.
If one denotes by $d_H$ the fractal (Hausdorff) dimension of Wiener paths contributing to the Feynman path integral, then in the Hartree-Fock approximation discussed here an upper critical dimension
of six would imply $d_H =3$, if the arguments in \cite{aiz81,aiz85,fro82} are followed here as well.
On the other hand, for a Gaussian scalar field (a free particle with no spin)
it is known that $d_H=2 $, as for regular Brownian motion \cite{fey65,kle06,zin05}.

A few words should be spent here on the physical interpretation of $m$ in gauge theories.
When one refers to the mass gap, it means a gap in energy between the ground state
and the first excited stated, $m = E_1 - E_0 $, as derived from the quantum
Hamiltonian (or transfer matrix) describing, in this case, the nonlinear sigma model. 
It's relationship with the Euclidean correlation length $\xi$ is provided by the Lehman
representation (i.e. completeness) for the two point function (as an example),
which then gives $m=1/ \xi$.
In addition, $\xi$ and thus $m$ are related by scaling to a multitude of other
observables, such as the order parameter or magnetization 
$\langle \sigma \rangle$ in the low temperature
phase for spin system, and the gluon condensate $\langle F_{\mu\nu}^2 \rangle$ 
for gauge theories.
In the gauge theory case one has a fundamental relationship between 
the nonperturbative scale $ \xi = m^{-1} $ and a nonvanishing vacuum expectation 
value for the gluon field \cite{cam89,xji95}.
\beq
\langle \, F_{\mu\nu}^2 \, \rangle  \;  \simeq \;  { 1 \over \xi^4 } \; .
\label{eq:vev_qcd}
\eeq
In QCD this last result is obtained from purely
dimensional arguments, once the existence of a fundamental
correlation length $\xi$ (inversely related to the mass gap) is established.
Actual physical values for the QCD condensates are well known;
current lattice and phenomenological estimates cluster around
$ \langle \, { \alpha_S \over \pi } \, F_{\mu\nu}^2 \, \rangle  \;
\simeq \;  (440 \, \text{MeV})^4 $ \cite{bro09,dom14}.
In gauge theories an additional physically related quantity 
is provided by the quark field condensate
\beq
\langle \, \bar \psi \psi \, \rangle \, \simeq \, { 1 \over \xi^3 } \; \; 
\label{eq:vev_fer}
\eeq
whose physical value is estimated at 
$ \langle \, \bar \psi \psi \, \rangle \, \simeq \, (280 \, \text{MeV})^3 $
\cite{mcn13}.
Again, the power of $\xi$ here is fixed by the canonical
dimension of the corresponding fermion field.

A possible physical explicit value for the mass gap parameter $m$ in $N$=3 QCD is provided by
the $J^{PC}=0^{++}$ spin zero 500 MeV $\sigma $ or $f_0 (500)$ glueball state, 
a very broad resonance seen in $\pi \pi $ and $\gamma \gamma$ scattering \cite{men08}.
If this particle is roughly considered as a spin-zero bound state of two (effectively massive)
spin-one gluons with antiparallel spins, then that would give for the mass gap
parameter $m \approx 250 \, \text{MeV} $.
For the scaling violation parameter on the r.h.s. of Eq.~(\ref{eq:m_lambda}) one then obtains
a value $ e^{\half} = 1.649 $ larger, or around $410 \, \text{MeV}$.
The latter is not too far off from the experimentally well-established value of 
$ \Lambda_{\overline{MS}} \approx 330 \, \text{MeV}$ for QCD (with three light
quark flavors).
Conversely, if one uses the current world average of 
$ \Lambda^{(3)}_{\overline{MS}} \approx (332  \pm 17) \, \text{MeV}$ 
\cite{pdg} for three flavor QCD, then one obtains for the mass of the 
$\sigma$ / $f_0$ meson $ m_{\sigma} \approx 403 \, \text{MeV} $.
These results seem to suggest that quantitatively the lowest order Hartree-Fock 
result is not expected to be better than a factor or two or so.
Many of these considerations will be useful later when the gravity case is discussed, 
due to the many deep analogies between gauge theories and quantum gravity.


\vskip 20pt

\section{The Quantum Gravity Case}
\label{sec:grav}

\vskip 10pt

Upon gaining confidence in this approach and technique, 
one can perform again a similar type of analysis in the case of quantum gravity, 
as outlined in detail in the preceding sections.  
In the end, one major result which makes the calculation feasible is the reliance
on well-known results for simple one loop diagrams.
This important insight makes it possible to avoid lengthy and complex one-loop 
gravity calculations, and rely instead,
as in all the previous cases, on suitably modified known diagrams via the insertion of an 
effective, dynamically generated mass parameter $m$.
The latter is then determined self-consistently via the solution of the 
resulting nonlinear gap equation, as discussed in detail for the nonlinear sigma
model (in Section \ref{sec:sigma_hf}), and for gauge theories
(in Section \ref{sec:gauge}).

\begin{figure}
	\begin{center}
		\includegraphics[width=0.75\textwidth]{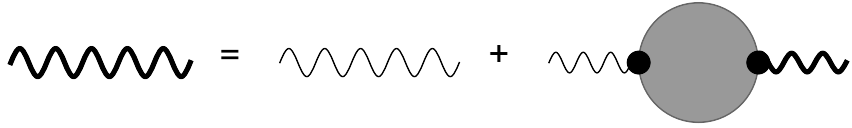}
	\end{center}
	\caption{Dressed graviton propagator $ D_{\alpha\beta,\mu\nu} (p)$, with a graviton proper vacuum polarization insertion $\Pi_{\alpha\beta,\mu\nu} (p)$. 
The thin wavy line denotes the tree-level graviton propagator $ D^0_{\alpha\beta,\mu\nu} (p)$.}
	\label{fig:gravity_proper}
\end{figure}

A number of new ingredients arise in the quantum gravity calculation, which
we proceed to enumerate here.
The first component is the (gauge choice dependent) graviton propagator 
\beq
D_{\alpha\beta,\mu\nu} (p) \;\; ,
\eeq
with the above expression denoting in general the dressed propagator, and 
$D^0_{\alpha\beta, \mu\nu} (p) $ the bare (tree-level) one.
An explicit form for the tree-level graviton propagator in covariant gauges, as well as the three-graviton 
and four-graviton vertex, can be found in the Feynman rules section of \cite{book}, 
and references therein, and for brevity will not be reproduced here.
Occasionally, in the following it will be convenient to suppress Lorentz indices altogether
in order to avoid unnecessary extensive cluttering.
Via loop corrections, the tree-level graviton propagator is then modified by gauge and matter vacuum 
polarization contributions $\Pi_{\mu\nu,\rho\sigma} (p) $, with the latter written as
\beq
\Pi_{\alpha\beta,\mu\nu} (p) \; = \; 
( p^2 \, \eta_{\alpha\beta}  - p_\alpha \, p_\beta ) \, 
( p^2 \, \eta_{\mu\nu} - p_\mu \, p_\nu ) \, \Pi (p^2)
\eeq
in terms of the scalar quantity $\Pi (p^2)$.
Here $\eta$ denotes the flat spacetime metric. 
The vacuum polarization contribution in quantum gravity is shown pictorially in
Figure \ref{fig:gravity_proper}.
By virtue of energy-momentum conservation, one then has the transversality
condition 
$p_\alpha \, \Pi^{\alpha\beta,\mu\nu} (p) = p_\mu \,  \Pi^{\alpha\beta,\mu\nu} (p) = 0 $,
and the graviton propagator (in the harmonic gauge \cite{vel75,book}) can be written as 
\beq
D_{\alpha\beta,\mu\nu} (p) \; = \; { I_{\alpha\beta,\mu\nu} \over p^2 \, [ \, 1 + \Pi (p^2) \, ] } \;\; 
\label{eq:grav_prop}
\eeq
with
$ I_{\alpha\beta,\mu\nu} = \eta_{\alpha\mu}  \eta_{\beta\nu} + \eta_{\alpha\nu}  \eta_{\beta\mu} -
\eta_{\mu\nu} \eta_{\alpha\beta} $.
As a consequence of gauge invariance (here more properly described
as general coordinate, or diffeomorphism, invariance) 
in perturbation theory, and to all orders, the graviton is expected to stay massless,
$ p^2 \, \Pi (p^2) \; \mathrel{\mathop\sim_{ p \rightarrow 0 }} \,  0 $.
Here we will find it quite useful that an explicit form for the vacuum polarization contribution 
$\Pi_{\alpha\beta,\mu\nu}  (p) $ in the context of perturbative
quantum gravity was given some time ago in \cite{cap73}, 
and such an explicit form will be used below.


\begin{figure}
	\begin{center}
		\includegraphics[width=0.50\textwidth]{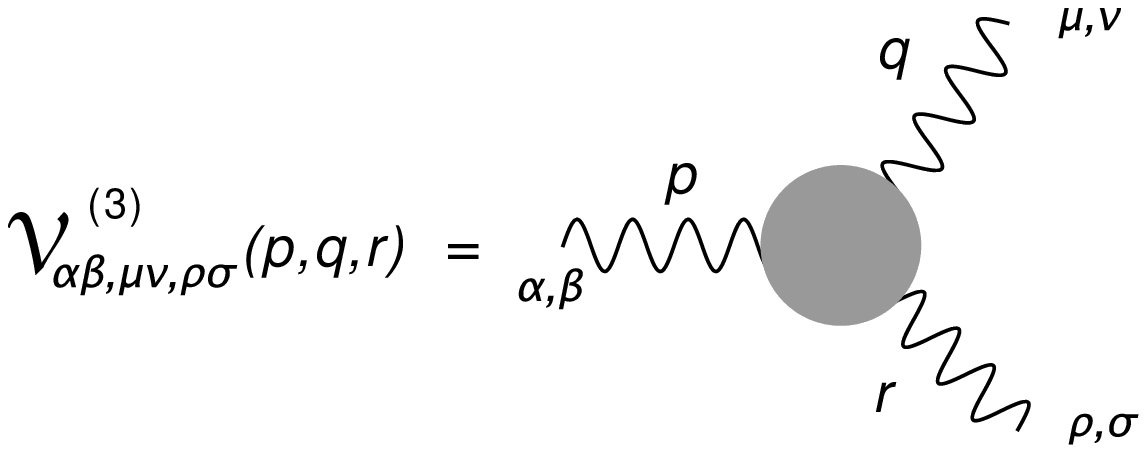}
	\end{center}
	\caption{Graviton three-point vertex ${V}^{(3)}_{\alpha\beta,\mu\nu,\rho\sigma} (p,q,r)
$.}
	\label{fig:gravity_3pt}
\end{figure}

\begin{figure}
	\begin{center}
		\includegraphics[width=0.50\textwidth]{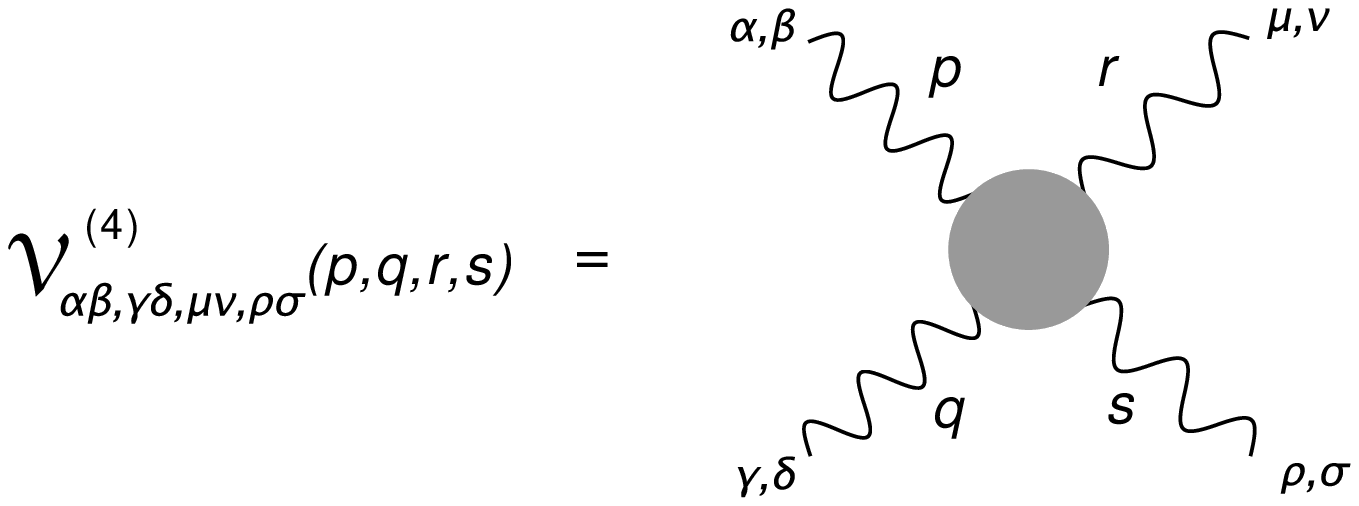}
	\end{center}
       \caption{Graviton four-point vertex ${V}^{(4)}_{\alpha\beta,\gamma\delta,\mu\nu,\rho\sigma} (p,q,r,s)$.}
	\label{fig:gravity_4pt}
\end{figure}

In addition, one has the dressed three-graviton vertex (shown here in Figure \ref{fig:gravity_3pt}) 
\beq
{V}^{(3)}_{\alpha\beta,\mu\nu,\rho\sigma} (p,q,r)
\label{eq:3_graviton}
\eeq
and the dressed four-graviton vertex (shown here in Figure \ref{fig:gravity_4pt}) 
\beq
{V}^{(4)}_{\alpha\beta,\gamma\delta,\mu\nu,\rho\sigma} (p,q,r,s) \;\; ,
\label{eq:4_graviton}
\eeq
and similarly for the ${\cal O} ( h^n )$ higher order graviton vertices, as they arise
in the weak field expansion of the Einstein-Hilbert action
about flat space, $ g_{\mu\nu}= \eta_{\mu\nu} + h_{\mu\nu} $.
Recall that for the Einstein-Hilbert action $\int \sqrt{g} \, R $ written in terms of the
weak field metric field $ h_{\mu\nu} $ one has 
\beq
\sqrt{g} = 1 + \half h - \quarter h_{\mu\nu} h^{\mu\nu} + \eigth h^2 + O(h^3)
\eeq
with trace $h= h_{\;\;\mu}^\mu$, and consequently (up to total derivatives) one obtains
schematically
\beq
I_{EH} \, = \, { 1\over G } \, \int d^d x \, \sqrt{ 1+ h} \; \partial^2 \, h \, \sim \,
{ 1\over G } \, \int d^d x \, \left [ \,  \half
h \, \Box \, h   \, - \, \eigth \, h^2 \, \Box \, h \, + \, {\cal O} (h^4) \right ] \;\; .
\eeq
The latter shows the origin of the trilinear ($h^3$) vertex, with weight proportional to 
momenta squared, $ \sim k^2$ in Fourier space, as well as the appearance of the
(infinitely many) higher order momentum-dependent vertices $ {\cal O} ( h^n )$.


\begin{figure}
	\begin{center}
		\includegraphics[width=0.80\textwidth]{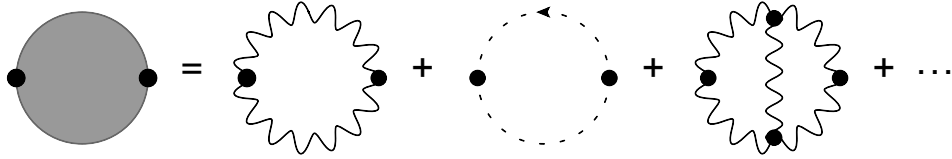}
	\end{center}
	\caption{Quantum gravity lowest order vacuum polarization $
\Pi_{\alpha\beta,\mu\nu} (p)$ diagrams. 
A thin wavy lines denotes a bare graviton propagator, and small sized dots stand for bare graviton vertices.
A dashed line appears in the ghost loop. Dots indicate higher order corrections.}
	\label{fig:gravity_vacpol}
\end{figure}

Two additional steps are needed in order to systematically develop the Hartree-Fock approximation for quantum gravity.
Firstly, and in close analogy to the previous discussion of the nonlinear sigma model
and gauge theories, one needs to write down Dyson's coupled equations for the graviton
propagator, vacuum polarization and proper vertices.
The lowest order Feynman diagrams (in a systematic perturbative expansion in Newton's 
constant $G$) as they contribute to the gravitational vacuum polarization are
shown, as an illustration, in Figure \ref{fig:gravity_vacpol}.
These generally contain contributions from graviton, matter and ghost loops.
To derive the full set of Dyson's equations one then follows the same procedure as in QED and
Yang-Mills theories, as outlined earlier in this paper in Section \ref{sec:hartree}.
First one writes down the full Feynman path integral (including gauge fixing and ghost terms), including
a classical source term $ g_{\mu\nu} (x) \, J^{\mu\nu} (x) $ for the gravitational field, as well as 
any additional fields present such as ghost and matter contributions.
One then applies suitable combinations of functional derivatives of the generating function
$Z[J]$ [as in Eqs.~(\ref{eq:qed_id}), (\ref{eq:qed_id1}) and (\ref{eq:qed_id2})] to obtain the full set of 
coupled integro-differential equations for the gravitational $n$-point functions, including the 
quantum version of Einstein's field equations [in close analogy to the QED result of Eqs.~(\ref{eq:max_quant}), (\ref{eq:max_quant1}) and (\ref{eq:max_quant2})].
As an example, Figure \ref{fig:gravity_proper} illustrates the relationship between the tree-level
graviton propagator to the dressed one via the proper gravity vacuum polarization contribution,
while Figure \ref{fig:gravity_vacpol} shows the lowest order Feynman diagram contributions to the vacuum polarization tensor for gravity.


\begin{figure}
	\begin{center}
		\includegraphics[width=0.75\textwidth]{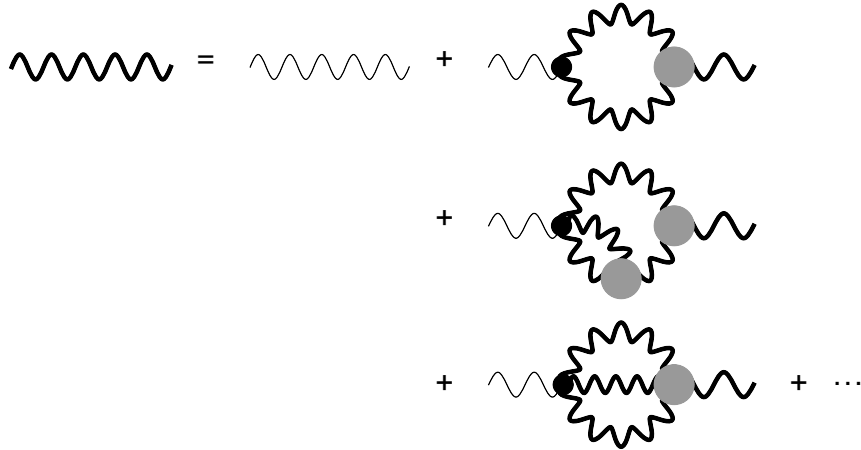}
	\end{center}
	\caption{Dyson's equations for the dressed graviton propagator $ D_{\alpha\beta,\mu\nu} (p)$. Small dots denote bare graviton vertices, while larger dots stand for dressed vertices.
Thin wavy lines denote bare graviton propagators, while thicker lines stand for dressed propagators.}
	\label{fig:gravity_dyson}
\end{figure} 


\begin{figure}
	\begin{center}
		\includegraphics[width=0.75\textwidth]{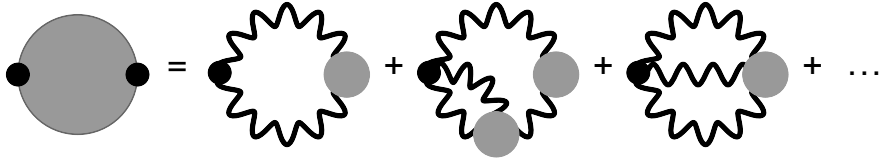}
	\end{center}
	\caption{Dyson's equations for the proper graviton vacuum polarization insertion 
$ \Pi_{\alpha\beta,\mu\nu} (p)$. 
Small dots denote bare graviton vertices, while larger dots stand for dressed vertices. 
A thin wavy lines denotes a bare graviton propagator, while thicker lines stand for dressed propagators.}
	\label{fig:gravity_dyson_self}
\end{figure} 

\begin{figure}
	\begin{center}
		\includegraphics[width=0.50\textwidth]{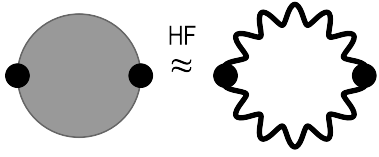}
	\end{center}
	\caption{Self-consistent Hartree-Fock approximation for the graviton proper vacuum polarization tensor. Small dots denote bare graviton vertices, and the thick wavy line indicates a dressed graviton propagator, to be determined self-consistently.}
	\label{fig:gravity_hf}
\end{figure}
\begin{figure}
	\begin{center}
		\includegraphics[width=0.80\textwidth]{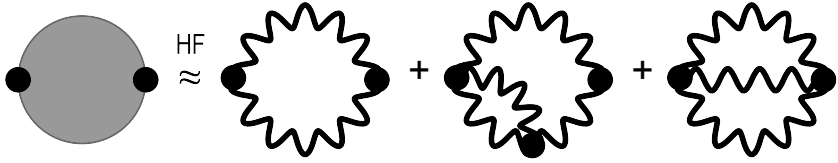}
	\end{center}
	\caption{Next higher order terms in the Hartree-Fock approximation to the graviton proper vacuum polarization tensor. Small dots denote bare graviton vertices, and the thick wavy line indicates a dressed graviton propagator, to be determined later self-consistently.}
	\label{fig:gravity_hf_higher}
\end{figure}

Next, we come to the actual form of Dyson's equations for quantum gravity.
Figure \ref{fig:gravity_dyson} shows Dyson's equation for the dressed graviton propagator (as written out in terms of the bare propagator and bare, as well as dressed, vertices), while
Figure \ref{fig:gravity_dyson_self} illustrates Dyson's equation for the proper gravity vacuum polarization term.
Here we give explicitly Dyson's equation for the dressed graviton propagator $D (p) $, in terms
of the bare propagator $D^0 (p)$ and the trilinear bare $V^{(3),0}$ and dressed $V^{(3)}$, as well as
the quadrilinear $V^{(4),0}$ and $V^{(4)}$ vertices, with the dots indicating contributions from
higher order vertices $V^{(n),0}$ and $V^{(n)}$ with $n>4$.
Written out explicitly, the equation reads
\begin{equation}
\begin{aligned}
D (p) & = D^0(p) \\
& + \frac{1}{2} \;\; D^0 (p) \cdot   \int\frac{d^dq}{(2\pi)^d}  \; 
V^{(3),0}(p,q,-p-q) \cdot D(p+q) \cdot 
V^{(3)}(p+q,-q,p) \cdot D(q) \cdot D(p) 
\\
&+ \frac{1}{3!} \; D^0 (p) \cdot \int\frac{d^dq_1 d^dq_2}{(2\pi)^{2d}}   \;  
V^{(4),0}(p,-q_1,q_2,-p-q_2-q_1) \cdot D(p+q_2-q_1) 
\\ 
& \;\;\;\;\;\;\;\;\;\;\;\;\;\;\;\;\;\,\, 
\cdot V^{(3)}(p+q_2-q_1,q_1-q_2,-p) \cdot D(q_1-q_2) \cdot  
V^{(3)}(q_1,-q_2,-q_1+q_2) \cdot D(q_2) 
\\
& \;\;\;\;\;\;\;\;\;\;\;\;\;\;\;\;\;\,\, 
\cdot D(q_1) \cdot D(p) 
\\
&+ \frac{1}{3!} \; D^0 (p) \cdot  \int\frac{d^dq_1 d^dq_2}{(2\pi)^{2d}}  \;  
V^{(4),0}(p,-q_2,q_1,-p-q_1+q_2) \cdot D(p+q_1-q_2) 
\\
& \;\;\;\;\;\;\;\;\;\;\;\;\;\;\;\;\;\,\, 
\cdot V^{(4)}(-q_1,q_2,-p,p+q_1-q_2) \cdot D(q_1) \cdot D(q_2) \cdot D(p) 
\\
& + \; \cdots
\end{aligned}
\label{eq:dse_gravity}
\end{equation}
The next step is the derivation of the Hartree-Fock approximation for quantum gravity.
As should be clear from the various cases discussed earlier, namely 
the nonlinear sigma model  (discussed in Section \ref{sec:sigma_hf}) and gauge theories
(discussed in Section \ref{sec:gauge}), one needs to focus on just the lowest
order loop diagrams.
Nevertheless, these will have bare graviton propagators replaced by dressed
ones, later to be determined self-consistently.
Figure \ref{fig:gravity_hf} illustrates the Hartree-Fock approximation for gravity, carried
out to the lowest order by just considering the lowest order graviton loop diagram, 
the subject of the current investigation.
On the other hand, figure \ref{fig:gravity_hf_higher} illustrates the next order Hartree-Fock approximation for gravity, which would include both the lowest order graviton loop diagram, 
as well as the next order graviton two-loop diagrams.
The latter will not be considered further here, but could in the future provide an improved answer, 
as well as useful quantitative estimates for the overall uncertainty.


In practice, for quantum gravity the relevant one loop integral for the vacuum polarization has the form
\beq
\int { d^d k \over ( 2 \pi )^d }  \,
{ V (p,-k-p,k ) \cdot V (-p,-k,k+p) \over k^2 \, (k+p)^2 } \;\; ,
\eeq
where the dot here indicates a generalized dot product over the relevant multitude of Lorentz
indices associated with the three-graviton, matter and ghost vertices.
Additional ingredients involve a symmetry factor $\half$ for the diagram, and of course the
gravitational coupling $G$.
In spite of the complexity of the one-loop expression for the graviton vacuum polarization
contribution, the actual integral that needs to be evaluated here has a rather simple form.
In the Hartree-Fock approximation, and in close analogy to the treatment
of the nonlinear sigma model and gauge theories described previously, 
one is generally left with an evaluation of the following integral,
\beq
S_d \; \int_0^\Lambda dk \, k^{d-1} \, { k^{ 2 \sigma}  \over (k^2 + m^2)^2 }
\label{eq:int_grav}
\eeq
with a combined spacetime volume factor $ S_d = 1 / 2^{d-1} \, \pi^{d/2}  \, \Gamma ( d/2 ) $, but 
now with $\sigma = 2$, specific to the gravitational case.
The latter steep momentum dependence is of course at the root of the perturbative
nonrenormalizability of quantum gravity in four dimension [15-29] :
the significant weight of the high momentum region in loop integrals renders
perturbation theory in Newton's $G$ badly divergent.

Going back to the required integral of Eq.~(\ref{eq:int_grav}), one finds for the momentum 
integral itself, in general dimension $d$,
\beq
\int_0^\Lambda dk \, k^{d-1} \, { k^{2 \, \sigma}  \over (k^2 + m^2)^2 }  =
{  \Lambda^{d+ 2} \, \sigma \over 2 \, m^2 \, (m^2 + \Lambda^2 )  } -
{ d + 2 \, \sigma - 2 \over d + 2 \, \sigma } \,
{  \Lambda^{d+ 2 \, \sigma}  \over 2 \, m^4  } \,
{}_2 F_1 ( 1, {d \over 2} +\sigma,  1 + { d \over 2} + \sigma, - { \Lambda^2 \over m^2 } ) \;\; ,
\label{eq:integral_grav}
\eeq
with ${}_2 F_1 ( a,b,c,z) $ the generalized hypergeometric function.
Here again $\Lambda$ is the ultraviolet cutoff, such as one implemented
in the Regge-Wheeler lattice formulation of gravity \cite{reg61,whe63}, for which 
$\Lambda \sim \pi / a $, with $a$ defined as an average lattice spacing 
$ a^2 \equiv \, \langle l^2 \rangle$. 
Indeed, as in all the previous examples, an explicit ultraviolet cutoff is required in order to
be able to write down a meaningful form of the Hartree-Fock equation.
Generally, for any quantum mechanical system a lattice cutoff is required to make the 
Feynman path integral well defined, as discussed already in great detail some time ago
by the inventor himself \cite{fey65}, and in some more recent monographs \cite{kle06,zin05}.
The root cause of this situation is that the dominant quantum paths contributing to the 
path integral are in general nowhere differentiable, a
key realization that motivated Regge and Wheeler to develop a lattice formulation for gravity\cite{reg61,whe63}.
For more details on the Regge-Wheeler lattice theory of gravity see \cite{book}, and further 
references therein.
The use of an explicit momentum cutoff leads to some subtle differences in the 
treatment of ultraviolet divergences, some of which are ordinarily sanctioned to be zero 
by the formal rules of dimensional regularization, such as
\beq
\int { d^d k \over ( 2 \, \pi )^d } \, { 1 \over k^n } \; = \; 0 \;\; .
\eeq
These nevertheless can be non-zero (and possibly highly divergent) when evaluated explicitly 
with both an infrared ($\mu$) and an ultraviolet ($\Lambda$) cutoff \cite{tho74,book}.

In quantum gravity another subtlety arises from the fact that one can easily reabsorb some 
troublesome ultraviolet divergences by a simple rescaling of the metric \cite{book}.
Consider the metric field redefinition
\beq
g_{\mu\nu} = \omega \, g_{\mu\nu}'
\label{eq:metric_scale}
\eeq
with $\omega$ a constant.
For pure gravity with a cosmological constant term proportional to $\lambda$, one writes for 
the Lagrangean
\beq
{\cal L} = - { 1 \over 16 \pi G} \, \sqrt{g} \, R \, + \, \lambda \sqrt{g}   \;\; .
\eeq
Under a rescaling of the metric as in Eq.~(\ref{eq:metric_scale}), one obtains
\beq
{\cal L} = - { 1 \over 16 \pi G} \, \omega^{d/2-1} \, \sqrt{g'} \, R' 
\, + \, \lambda \, \omega^{d/2} \, \sqrt{g'} \;\; ,
\label{eq:rescale}
\eeq
which just amounts to a rescaling of Newton's constant and the cosmological constant
\beq
G \rightarrow \omega^{-d/2+1} G \; , \;\;\;\; 
\lambda \rightarrow \lambda \, \omega^{d/2}  \;\; ,
\eeq
leaving the dimensionless combination $G^d \lambda^{d-2}$ unchanged.
As a consequence, it seems physically meaningless to discuss separately the renormalization properties
of $G$ and $\lambda$, as they are both individually gauge-dependent
in the sense just illustrated.
To some extent these arguments should contribute to clarifying
why in the following it will be sufficient to focus on the renormalization properties of 
Newton's constant $G$.
In addition, one issue we have not touched here at all is the role played by the gravitational
functional measure (normally taken to be the De Witt one, see \cite{book} 
and references therein) in cancelling spurious divergences. 
The issue is a rather subtle one, whose importance was already emphasized in the earlier
discussion of the non-linear sigma model, where the role of the functional measure in canceling
spurious tadpole divergences was pointed out in Eq.~(\ref{eq:jacobian}).
In the gravity case, the important role of the gravitational functional measure contribution 
in canceling spurious ultraviolet divergences can be seen from the following simple 
argument given in \cite{book}.
Consider the following covariant (DeWitt) gravitational measure
\beq
\int \, \left [ d \, g_{\mu\nu} \right ] \, \equiv \, 
\int \, \prod_x \, \left [ g(x) \right ]^\sigma \, \prod_{\mu \ge \nu} \, d \, g_{\mu\nu} (x) 
\label{eq:grav_meas}
\eeq
with $\sigma$ a real (positive or negative) parameter.
Then one has
\beq
\prod_x \, \left [ g(x) \right ]^\sigma \, \prod_{\mu \ge \nu} \, d \, g_{\mu\nu} (x) 
\, = \, \left ( \prod_x \, \prod_{\mu \ge \nu} \, d \, g_{\mu\nu} (x) \right ) \;
\exp \left \{ \sigma \, \delta^d (0) \, \int d^d x \, \log g(x) \right \} \;\; .
\label{eq:grav_meas1}
\eeq
Here, given a proper lattice regularization (such as the Regge-Wheeler one) with ultraviolet cutoff 
$\Lambda$, one has $ \delta^d (0) = ( \Lambda / \pi )^d $.
Then the above argument shows that a proper choice of functional measure can play an essential role
in canceling spurious $\Lambda^d $ divergences in $d$ spacetime dimensions.

From the integral in Eq.~(\ref{eq:integral_grav}) one then obtains the general nonlinear
gap equation for the parameter $m$ in $d$ dimensions, which reads
\beq
G \; \beta_0 \; S_d \, \left [ \,
{ d + 2 \over 2 \, (d + 4) } \;
{  \Lambda^{d+ 4 }  \over m^4  } \;
{}_2 F_1 ( 1, {d \over 2} + 2, { d \over 2} + 3 , - { \Lambda^2 \over m^2 } ) 
\, - \, { \Lambda^{d+ 2} \over m^2 \, (m^2 + \Lambda^2 )  } 
\right ] \; = \; 1 \;\; .
\label{eq:gap_eq_d}
\eeq
Here $G$ is Newton's constant, and $\beta_0$ an overall numerical coefficient 
resulting from Lorentz index traces, diagram multiplicities and individual
diagram symmetry factors.
In the above equation $G$ is dimensionful, with mass dimension $d-2$, 
$G \sim \Lambda^{2-d}$.
The next step is to discuss the solution to this nonlinear equation in various dimensions,
both to draw parallels to the nonlinear sigma model and gauge theory cases
outlined earlier, and to explore the sensitivity of the results to the dimension $d$.


It will be useful here to digress very briefly about the physics of quantum gravity in
$ d < 4 $.
In general it is possible at least in principle to define quantum gravity
in any $d>2$.
There are $d(d+1)/2$ independent components of the
metric in $d$ dimensions, and the same number of algebraically independent
components of the Ricci tensor appearing in the field equations. 
The contracted Bianchi identities reduce the
count by $d$, and so does general coordinate invariance,
leaving $d(d-3)/2$ physical gravitational
degrees of freedom in $d$ dimensions.
At the same time, four space-time dimensions is known to be the lowest dimension
for which Ricci flatness does not imply the vanishing of the gravitational
field, $R_{\mu\nu\lambda\sigma}=0$, and therefore the first dimension
to allow for gravitational waves and their quantum counterparts, gravitons.
Note that in a general dimension $d$, the position space tree-level graviton propagator obtained from 
the linearized theory, given earlier in $k$-space in Eq.~(\ref{eq:grav_prop}),
is obtained by Fourier transform.
It is proportional to the integral
\beq
\int d^d k \, { 1 \over k^2 } \, e^{i \, k \cdot x } \; = \;
{ \Gamma \left ( {d - 2 \over 2} \right ) \over 4 \, \pi^{d/2} \,
( x^2 )^ { d /2 - 1} } \;\; ,
\eeq
while the static gravitational potential involves just the spatial Fourier transform
\beq
V(r) \, \propto \, 
\int d^{d-1} {\bf k} \, { e^{i {\bf k} \cdot {\bf x} } \over {\bf k}^2 }
\, \sim \, { 1 \over r^{d-3} } \;\; .
\eeq

In two spacetime dimensions there are no genuine gravitational
(transverse-traceless) degrees of freedom,
the Einstein-Hilbert action gives a topological invariant proportional to the Euler characteristic
of the manifold.
The only surviving degree of freedom is the conformal mode, which in two dimensions enters the 
gravitational action as a total derivative.
In the absence of matter, the only residual interaction is then generated by
the non-local Fadeev-Popov determinant associated with the choice of conformal gauge.
Nevertheless it is possible to consider, as was done some time ago in \cite{gas78,chr78,wei79} and
later in [90-95], two dimensions as the $\epsilon \equiv d-2 \rightarrow 0 $ limit for 
gravity formulated in $2+\epsilon$ dimensions, with $\epsilon$ considered a small quantity.
This followed a similar procedure, the dimensional expansion, pioneered earlier by
Wilson for scalar fields \cite{wil72,wil72a,wil73}.
It was also realized early on that in the case of gravity this limit is rather subtle, due to confluent
kinematic singularities as $d \rightarrow 2 $.

In two spacetime dimensions ($d=2$) the gravitational Hartree-Fock gap equation of Eq.~(\ref{eq:gap_eq_d}) reduces to
\beq
{ G \, \beta_0 \over 2 \pi } \, \left [ \, 2 \, 
\log \left (  { \Lambda \over m } \right ) \, - \half \, \right ]
\; = \; 1  \;\; .
\label{eq:gap_eq_2d}
\eeq
A spurious quadratic divergence in the integral given above needs to be
subtracted, in order to achieve consistency with the known, more recent perturbative
result for gravity in $2+\epsilon$ dimensions given in [91-95], 
discussed in greater detail further below.
The critical point (or ultraviolet fixed point) in this case is located at the origin
\beq
G_c \; = \; 0  \;\; .
\label{eq:gc_2d}
\eeq
The solution to the above $d=2$ gap equation then has the form
\beq
m \; = \; \Lambda \, e^{- \quarter } \; e^{- { \pi  \over \beta_0 \, G } } 
\; = \; \Lambda \, e^{- \quarter} \; e^{- { 3 \pi  \over 25 \, G } }  \;\; ,
\label{eq:massgap_2d}
\eeq
where at the end the correct expression $\beta_0  = 25/3 $ for pure gravity  in
$d=2$ derived in \cite{eps4} has been inserted.
Here $G \equiv G ( \Lambda ) $ is the bare coupling at the cutoff scale $\Lambda$.
The above results then implies, within the current approximation,
that the dynamically induced gravitational mass gap $m$ is by a factor 
$ e^{-\quarter} = 0.7788 $ smaller than the gravitational scaling parameter 
$ \Lambda \, \exp ( - \pi / \beta_0 \, G ) $.

The above result then leads to some rather immediate consequences, again very 
much in line with the previous discussion for scalar fields in Section \ref{sec:sigma_hf}, 
and for gauge theories in Section \ref{sec:gauge}.
Thus the renormalization group running of the gravitational coupling $G(\mu)$ 
follows again from the requirement that the mass gap parameter $m$ 
be scale independent,
\beq
\mu \, { d \, m \over d \, \mu } \; = \;  0 \; = \;
\mu \, { d \, \over d \, \mu } \, \left ( 
\mu \, e^{- \quarter} \, e^{- { \pi \over \beta_0 \, G ( \mu ) } }  \right )  \; .
\eeq
This then gives the expected asymptotic freedom-like running of the gravitational 
coupling $G(\mu)$ in the vicinity of the $d=2$ ultraviolet fixed point located at $G=0$,
namely
\beq
G ( \mu ) \; = \;  { \pi \over  \beta_0 \, [ \, \log ( \mu /  m )  - \quarter  \, ] }  \;\; .
\label{eq:grun_2d}
\eeq
One notes here that the quantity $m$ plays a role quite analogous to the $\Lambda_{\bar MS} $ of
QCD and, more generally, $SU(N)$ gauge theories, as discussed here earlier in Section \ref{sec:gauge}.
Indeed, the quantum gravity result in $d=2$ looks rather similar to the 
result for the two-dimensional nonlinear sigma model (for $N>2$) and to the gauge theory
result in $d=4$, in the sense that all three models have a mass gap that
vanishes exponentially at zero coupling.
This last fact is of course already well known from the simplest application of 
perturbation theory.
Nevertheless, the gravity result in two dimensions can be regarded as of
very limited physical interest, given that the Einstein-Hilbert action is just a topological
invariant in $d=2$.
It is also worthy of note here that, overall, the main results follow from the structure of the
one loop integral - and from the sign of $\beta_0$ - but are otherwise universal,
in the sense that a specific value for the magnitude of $\beta_0$ does not
play much of a role in the scaling laws, in the sense that quite generally
the mass gap has an essential singularity at $G=0$ in $d=2$.


Before proceeding to discuss the Hartree-Fock results in higher dimensions, it will be
useful here to recall the main results of the $2+\epsilon$ expansion for gravity [91-95].
To first order in $G$ one finds
\beq
\mu { \partial \over \partial \mu } \, G \, \equiv \, 
\beta (G) = \epsilon \, G \, - \, \beta_0 \, G^2 \, + \,O( G^3, G^2 \epsilon )
\label{eq:beta_oneloop}
\eeq 
with $\beta_0 = \twoth \cdot (25-c)$ where $c$ is the central charge for the 
massless matter fields.
Later the original one-loop calculation was laboriously extended 
to two loops \cite{eps4}, with the result
\beq
\mu { \partial \over \partial \mu } G \, \equiv \,
\beta (G) = \epsilon \, G \, - \, { 2 \over 3 } \, (25 - c) \, G^2 \,
- \, { 20 \over 3 } \, (25 - c) \, G^3 \, + \dots \;\; ,
\label{eq:beta_twoloops}
\eeq
again for $c$ massless real scalar fields minimally coupled to gravity.
At the nontrivial ultraviolet fixed point, for which $\beta (G_c) \, = \, 0 $,
\beq
G_c \, = \,  {3 \over 2 \, ( 25 - c ) } \, \epsilon
\, - \, {45 \over 2 ( 25 - c )^2 } \, \epsilon^2 \, + \, \dots
\label{eq:gceps}
\eeq
the theory becomes scale invariant by definition.
In statistical field theory language, the fixed point corresponds to a phase transition
where the correlation length $\xi=1/m$ diverges in units of the cutoff.
If one defines the correlation length exponent $\nu$ in the usual way by
\beq
\beta ' (G_c) \, = \, - 1/ \nu
\label{eq:nu_beta}
\eeq
then one finds the expansion in $\epsilon$
\beq
\nu^{-1} \, = \,  \epsilon \, + \, {15 \over 25 - c } \, \epsilon^2 \, + \, \dots
\label{eq:nueps}
\eeq
This then gives, for pure gravity without matter ($c=0$) in four dimensions (d=4), to lowest order
$\nu^{-1} = 2$, and $\nu^{-1} \approx 4.4 $ at the next order.
One can integrate the $\beta$-function equation 
in Eq.~(\ref{eq:beta_twoloops}) to obtain the renormalization
group invariant quantity
\beq
m \, = \, \xi^{-1} \, = \,  {\rm const.} \; \Lambda \, \exp \left ( - \int^G { dG' \over \beta (G') } \right )
\label{eq:xi_beta_grav}
\eeq 
which is identified with the mass gap $m$, or equivalently the inverse correlation length $\xi$.
The multiplicative constant in front of the last expression on the right
hand side arises as an integration constant of the renormalization group equation, 
and thus cannot be determined from perturbation theory in $G$ alone.
From Eq.~(\ref{eq:beta_twoloops}), in the $2+\epsilon$ expansion the running of Newton's
$G$ in the strong coupling (antiscreening) phase with $\epsilon > 0$ and $G>G_c$ 
is given by
\beq
G( \mu ) \; \simeq \; G_c \, \left [ 
\, 1 \, + \, c_0 \, \left ( {m^2 \over \mu^2 } \right )^{(d-2)/2}
\, + \, \dots \right ]
\label{eq:grun_cont} 
\eeq
where the dots indicate higher order radiative corrections.
Again, it is noteworthy here that the quantum amplitude $c_0$ arises as an integration constant of
the renormalization group equations, and thus (unlike the Hartree-Fock result)
remains undetermined in perturbation theory.


Next consider the case of quantum gravity in a spacetime dimension $d=3$.
Even in three dimensions it is known that there are no genuine gravitational
(transverse-traceless) degrees of freedom, with the only dynamical surviving 
remnant being the conformal mode \cite{des84}.
In $d=3$ the gravitational gap equation of Eq.~(\ref{eq:gap_eq_d}) for small $m$ reduces to
\beq
{ G \, \beta_0 \over 2 \pi^2 } \, \left [ \, 
2 \, \Lambda \, - \, { 5 \, \pi \over 4 } \, m \, - \, 
{ 3 m^2 \over \Lambda } \, \right ] \; = \; 1  \;\; .
\label{eq:gap_eq_3d}
\eeq
Again, a spurious cubic divergence in the integral needs to be subtracted, as discussed
earlier, in order to achieve consistency with the known perturbative
results for gravity in $2+\epsilon$ dimensions.
The solution to the above equation then has the form
\beq
m \; = \; \Lambda \, { 8 \over 5 \, G_c } \, \left ( G - G_c \right ) 
\, + \, {\cal O} \left (  (G-G_c)^2  \right ) \;\; .
\label{eq:massgap_3d}
\eeq
What is new here is the appearance of a critical point in $G$, located at
\beq
G_c \, = \, { \pi^2 \over \beta_0 \, \Lambda } \;\; ,
\label{eq:gc_3d}
\eeq
with a mass gap exponent [the power on the r.h.s. of Eq.~(\ref{eq:massgap_3d})] $\nu = 1 $.
This last result is seen here to be universal, in the sense that it does not depend
on the magnitude of $\beta_0$, $\Lambda$ or $G_c$.
One can roughly estimate the magnitude of $\beta_0$ in $d=3$ by interpolating 
between the value in $d=2$ and the value in $d=4$ given later below.
By this procedure one obtains approximately $\beta_0 \simeq (25+47)/6 = 12 $
and thus $G_c \simeq 0.822 / \Lambda $, not too far off
from the $d=3$ lattice gravity value $ G_c \approx 0.505 / \Lambda $ given some
time ago in \cite{hw93}.

The relationship between $m$ and $G (\Lambda) $ in Eq.~(\ref{eq:massgap_3d}) can 
then be inverted, to produce the general renormalization group running of $G(\mu)$.
The latter follows, as before, from the requirement that the mass gap $m$ 
be scale independent $ \mu \, { d \, m \over d \, \mu } \; = \;  0 $.
One then finds for $G(\mu)$ in the vicinity of the $d=3$ gravity nontrivial
ultraviolet fixed point
\beq
G ( \mu ) \; = \;  G_c \left [  \, 1 \, + \, { 5 \, \pi \over 8 } \, { m \over \mu } \, + \,  
{\cal O} \left (  \left ( { m \over \mu } \right )^2  \right ) \right ] \;\; .
\label{eq:grun_3d}
\eeq
Note again that the coefficient $\beta_0$ does not affect the scaling directly;
it only enters indirectly through the value of $G_c$, which in any case is affected by
the choice of the ultraviolet cutoff $\Lambda$.
More importantly, the ultraviolet cutoff $\Lambda$ has disappeared entirely from the expression
of Eq.~(\ref{eq:grun_3d}) for $G (\mu)$  (it is hidden in the physical parameter $m$, and in $G_c$).
The main result here follows again, quite generally, from the structure of the
one loop integral - and from the sign of $\beta_0$ - but is otherwise universal,
in the sense that a specific value for the magnitude of $\beta_0$ does not
play much of a role in the scaling laws:
the mass gap in $d=3$ vanishes linearly at $G_c$ within the
Hartree-Fock approximation.
This last result, which implies a mass gap exponent $\nu=1$, can be compared to 
the numerical lattice gravity result in $d=3$ of \cite{hw93} which gave $\nu= 0.59 (2)$, 
as well as to the exact solution of the lattice Wheeler-DeWitt equation in $2+1$ dimension,
which gives $\nu = 2/3 $ \cite{htw12,htw13,ht20}.
Note again here the similarity with the corresponding results
for the $d=3$ nonlinear sigma model, and for gauge theories in $d=5$ discussed
earlier.


Next we come to the physical case of quantum gravity in four spacetime dimensions ($d=4$).
The gravitational gap equation of Eq.~(\ref{eq:gap_eq_d}) for small $m$ reduces to
\beq
{ G \, \beta_0 \over 8 \pi^2 } \, \left [ \, 
\Lambda^2 \, + \, \half \, m^2  \, - \, 3 \, m^2 \, \log \left ( { \Lambda \over m } \right ) 
\, \right ]  \; = \; 1  \;\; .
\label{eq:gap_eq_4d}
\eeq
Again, a spurious quartic divergence in the integral needs to be subtracted in order 
to achieve consistency with the known perturbative results for gravity in 
$2+\epsilon$ dimensions;
as mentioned previously, in the $2+\epsilon$ expansion the latter is removed by a suitable 
rescaling of the metric.
Furthermore, one would need to consider here the role played by the gravitational
functional measure (normally taken to be the De Witt one, see \cite{book} and 
references therein) in cancelling spurious $\delta (0) \simeq ( \Lambda / \pi )^d $ 
tadpole divergences, as in Eq.~(\ref{eq:grav_meas1}).
The measure's effect was already emphasized earlier in the discussion of the non-linear 
sigma model, where its role in canceling spurious tadpole divergences was pointed out 
in the discussion surrounding Eq.~(\ref{eq:jacobian}).

The solution to the above $d=4$ gap equation then has the form
\beq
m \; = \; \Lambda \, \sqrt{ 2 \over 3 \, G_c } \, 
\left \vert  G  - G_c \right \vert^{1 \over 2} \,
\left [ - \log \left \vert { G - G_c \over G_c } \right \vert \, \right ]^{- {1 \over 2} }
\, + \, {\cal O} \left (  G-G_c  \right ) \;\; .
\label{eq:massgap_4d}
\eeq
Note the appearance of a new ingredient, a logarithmic correction similar
to the nonlinear sigma model result in $d=4$, Eq.~(\ref{eq:mass_nonlin_4d}).
Here again $G \equiv G ( \Lambda ) $ is the bare coupling at the cutoff scale $\Lambda$.
In four dimensions one finds that the critical point in Newton's $G$ is located at
\beq
G_c \, = \, { 8 \, \pi^2 \over \beta_0 \, \Lambda^2 } \;\; ,
\label{eq:gc_4d}
\eeq
and for the mass gap one now has an exponent $\nu = 1/2 $.
This last result is again universal, in the sense that it does not depend
on the magnitude of $\beta_0$, nor on the value of $G_c$.

The above $d=4$ gravity results share some significant 
similarity with the corresponding 
results for the $d=4$ nonlinear sigma model, and for gauge theories in $d=6$.
Recall that the main result here follows generally from the structure of the
one loop integral - and from the sign of $\beta_0$ - but is otherwise universal,
in the sense that a specific value for the magnitude of $\beta_0$ does not
play any role in the scaling laws:
the mass gap in $d=4$ goes to zero with a square root singularity (up to log
corrections) at $G_c$, at least within the Hartree-Fock approximation.

For gravity one can use again the Nielsen-Hughes formula \cite{nie81,hug80} to 
provide an estimate for the numerical coefficient $\beta_0$ in $d=4$.
The one-loop $\beta$-function contribution coefficient 
arising from a particle of spin $s$ running around the loop is 
\beq
\beta \, = \, - (-1)^{2 s} \, [ (2 s)^2 - \third ]  \;\; .
\label{eq:nh}
\eeq
For spin $s=2$ the above formula then gives for pure gravity (no matter or radiation fields) in four
dimensions $\beta_0  = 47 / 3 $, quite a bit larger than the gauge theory case
(and with an overall negative sign in the beta function, 
correctly taken into account previously).
This then gives from Eq.~(\ref{eq:nh}) $G_c = 5.035 / \Lambda^2 $, compared to the 
lattice gravity value in $d=4$ $ G_c = 1.069 / \Lambda^2 $ given in \cite{ham15}.
A significant quantitative difference between the two results is not surprising, given
the entirely different nature of the ultraviolet cutoff in the two cases (momentum 
cutoff vs. space-time lattice cutoff).

The relationship between $m$ and $G(\Lambda)$ in Eq.~(\ref{eq:massgap_4d}) can 
be inverted to give the renormalization group running of Newton's $G(\mu)$, as was done
earlier, and it follows from the requirement that the mass gap $m$ 
be scale independent, $\mu \, { d \, m \over d \, \mu } \; = \;  0 $.
One then finds for $G(\mu)$ in the vicinity of the $d=4$ gravity nontrivial
renormalization group ultraviolet fixed point the explicit result
\beq
G ( \mu ) \; = \;  G_c \left [  \, 1 \, - \, 
{ 3 \over 2 } \, \left ( { m \over \mu } \right )^2 \,
\log \left ( { 3 \, m^2 \over 2 \, \mu^2 } \right )
\, + \, {\cal O} \left (  \left ( { m \over \mu } \right )^4  \; \right ) \right ] \;\; .
\label{eq:grun_4d}
\eeq
In principle there are here two solutions for $G > G_c$ and $G < G_c$, which differ by
the sign of the quantum correction.
Here we have chosen, in accordance with the results of the lattice calculations
of \cite{ham15}, the sign that corresponds to gravitational antiscreening, since
the weak coupling gravitational screening phase is nonperturbatively unstable.
It is noteworthy here that the coefficient $\beta_0$ does not enter renormalization
group running directly,
it only enters indirectly through the value of $m$ and $G_c$, and the 
latter is in any case affected by the choice of ultraviolet cutoff.
Physically the result of Eq.~(\ref{eq:grun_4d}) implies that for $d=4$ and
in real space the quantum correction to Newton's constant initially grows 
quadratically (up to the slowly varying log correction) with distance 
in the strong coupling phase, $ \delta G ( r ) \sim ( r / \xi )^2 $, with $\xi = 1/m $.
Perhaps one of the most interesting aspects of the above result is the value of the universal 
quantum amplitude $c_0 = 3/2 $ of the $m^2/\mu^2 $ correction, as well as
the weak $log$ correction appearing in four dimensions.
Note that the power of $m/ \mu $ appearing in the formula for $G(\mu)$  is given 
generally (by definition) by $1 / \nu $, so that here one has in the 
Hartree-Fock approximation a power $ \nu = 1/2 $.
Note that the quantity $G_c$ (the fixed point value for Newton's constant) can be entirely
eliminated if one compares the strength of the gravitational coupling on two different 
momentum scales $\mu_1$ and $\mu_2$,
\beq
{ G ( \mu_1 ) \over G ( \mu_2 ) } \; = \;  
{ 
\, 1 \, - \, { 3 \over 2 } \, \left ( { m \over \mu_1 } \right )^2 \,
\log \left ( { 3 \, m^2 \over 2 \, \mu_2^2 }  \right ) \, + \, \cdots
\over
\, 1 \, - \, { 3 \over 2 } \, \left ( { m \over \mu_2 } \right )^2 \,
\log \left ( { 3 \, m^2 \over 2 \, \mu_2^2 } \right )  \, + \, \cdots
}  \;\; ,
\label{eq:grun_4da}
\eeq
again in the vicinity of the ultraviolet fixed point, and thus up to higher order 
corrections in $ m / \mu $.

Next consider the universal scaling exponent $\nu$, which is $\nu= \half$ for the Hartree-Fock 
approximation in four dimensions, as can be seen from Eq.~(\ref{eq:massgap_4d}).
As a direct consequence, one finds a power of $1/\nu=2$ in the momentum dependence of 
$G(\mu)$ in Eq.~(\ref{eq:grun_4d}).
The value of $\nu$ can then be compared to the lattice gravity estimate, where
one finds in four dimensions $ \nu \simeq 1/3 $, which then implies a larger
power of $3$ for the running of $G(\mu)$ (as given below).
Indeed it is useful at this stage to compare the Hartree-Fock result for $G(\mu)$  to the 
corresponding result in the Regge-Wheeler lattice theory of gravity \cite{ham15}, 
where one finds for the running of Newton's $G$ 
\beq
G_{Latt} ( \mu ) \; = \;  G_c \left [  \, 1 \, + \, c_0 \, \left ( { m \over \mu } \right )^3
\, + \, {\cal O} \left (  \left ( { m \over \mu } \right )^6  \right ) \right ] \;\; .
\label{eq:grun_latt}
\eeq
One notes that the structure of the running is similar to the Hartree-Fock result, 
nevertheless the powers are different
(two versus three) and there is some difference in the numerical value of the amplitude as well.
The amplitude in the lattice case, $c_0 \simeq 16.0 $ \cite{ham15},
is thus somewhat larger (by a factor of five) than the Hartree-Fock value $ c_0= 3/2 $ given above.
A recent comparison of the running of $G$ with current cosmological observations, to be discussed later here, leads to useful observational constraints on the amplitude $c_0$, suggesting a value 
$ c_0 \approx 2.29 $ and thus smaller than the current lattice estimate, and more in line with the above Hartree-Fock result.


It pays to look briefly at a few cases of dimension greater than four.
In five spacetime dimensions ($d=5$) the gravitational gap equation of 
Eq.~(\ref{eq:gap_eq_d}) for small $m$ reduces to
\beq
{ G \, \beta_0 \over 12 \pi^3 } \, \left [ \, 
{ 2 \over 3 } \, \Lambda^3 \, - \, 3 \, m^2  \, \Lambda \, \right ]
\; = \; 1  \;\; ,
\label{eq:gap_eq_5d}
\eeq
again with the spurious leading $\Lambda^5 $ divergence subtracted in order 
to achieve consistency with the known perturbative result for gravity in $2+\epsilon$ dimensions.
The solution to the above equation then has the form
\beq
m \; = \; \Lambda \, \sqrt{ 2 \over 3 \, G_c } \, 
\sqrt{  G  - G_c } \, + \, {\cal O} \left (  G-G_c  \right ) \;\; .
\label{eq:massgap_5d}
\eeq
The critical point in Newton's $G$ is now located at
\beq
G_c \, = \, { 18 \, \pi^3 \over \beta_0 \, \Lambda^3 } \;\; .
\label{eq:gc_5d}
\eeq
For the mass gap exponent one has again the universal value $\nu = 1/2 $,
independent of $\beta_0$ or $G_c$.
As before, the above relation between $m$ and bare $G$ can be inverted to give the 
renormalization group running of $G(\mu)$, 
from $ \mu \, { d \, m \over d \, \mu } \; = \;  0 $.
This then gives for $G(\mu)$ in the vicinity of the $d=5$ gravity ultraviolet fixed point
\beq
G ( \mu ) \; = \;  G_c \left [  \, 1 \, + \, 
{ 9 \over 2 } \, \left ( { m \over \mu } \right )^2
\, + \, {\cal O} \left (  \left ( { m \over \mu } \right )^4  \right ) \right ] \;\; ,
\label{eq:grun_5d}
\eeq
with, in this case, no log correction.
As expected the power in the running is $1 / \nu =2 $.
In turn the result of Eq.~(\ref{eq:grun_5d}) implies that here 
in real space the quantum correction to Newton's constant initially grows quadratically 
with distance in the strong coupling phase, $ \delta G ( r ) \sim ( r / \xi )^2 $, with $\xi = 1/m $.
Here the quantum amplitude coefficient in $d=5$ is $c_0 = 9/2=4.5 $ in the Hartree-Fock approximation,
which is about one fourth the value found on the lattice in $d=4$, $c_0 \simeq 16.0 $, see \cite{vac17,ham15} and references therein.
We will return to this point later on.

Overall, the gravity results here share some similarity with the corresponding results
for the $d>4$ nonlinear sigma model, and for gauge theories in $d>6$.
They also suggest that the upper critical dimension for gravity is $ d_c = 4 $, as
in the case of scalar (spin zero) field theories.
Moreover, if one denotes by $d_H$ the fractal (Hausdorff) dimension of Wiener paths contributing 
to the Feynman path integral, then in the Hartree-Fock approximation an upper critical dimension
of four would imply $d_H =2$ if the arguments in \cite{aiz81,aiz85,fro82} are followed here as well.
In comparison, for a Gaussian scalar field (a free particle with no spin)
it has been known that $d_H=2 $, as for regular Brownian motion \cite{fey65,kle06,zin05}.
Indeed, a large-$d$ dimensional expansion was pursued for lattice gravity in \cite{spin,larged}.
Nevertheless, one finds that in this limit the lowest order results are not particularly illuminating
regarding to what might happen in four dimensions.
We should mention that a similar large-$d$ expansions were pursued for scalar field theories in 
\cite{eng63,fis64,abe72} and for gauge theories in \cite{dro79}, and in both cases the theory was
formulated on the lattice.
Some additional papers that also looked at quantum gravity in the continuum from the large $d$ point
of view can be found in \cite{str81,bjb04}.

\begin{figure} 
	\begin{center}
		\includegraphics[width=0.7\textwidth]{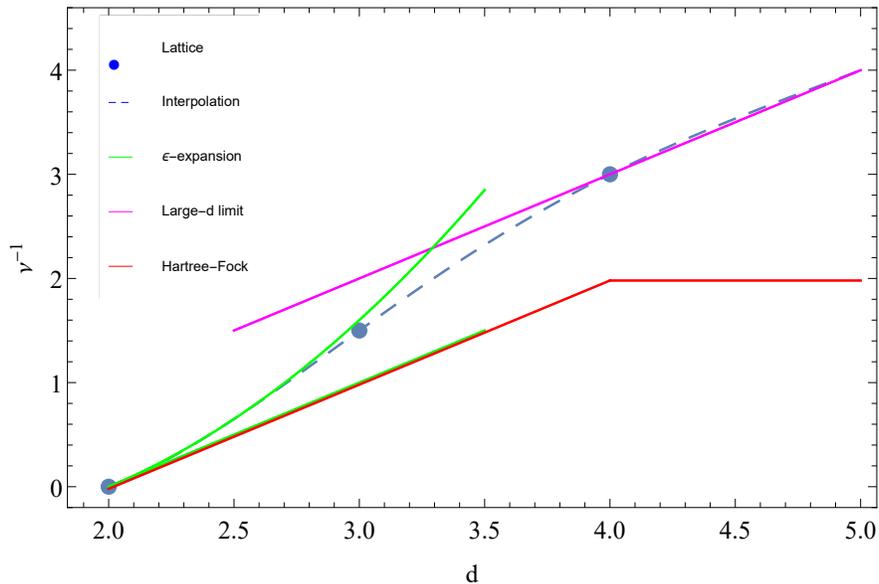}
	\end{center}
	\caption{
	Universal correlation length scaling exponent $\nu$ as a function of spacetime dimension $d$.
    	Shown are the $2+\epsilon$ expansion result to one (lower curve) and two 
       (upper curve) loops \cite{eps4},
	the value in $2+1$ dimensions obtained from the exact solution of the 
       Wheeler-DeWitt equation \cite{htw12,ht20}, 
       the numerical lattice result in four spacetime dimensions \cite{ham15}, 
	the large $d$ result $ \nu^{-1} \simeq d-1 $ \cite{spin}, and the Hartree-Fock
              approximation results discussed in the text.
	}
	\label{fig:exponents}
\end{figure}

It is possible to derive a general analytical expression for the mass gap $m$ in general dimensions.
In $d$ dimensions, with $d>4$, one finds from Eq.~(\ref{eq:gap_eq_d}) for the critical point
\beq
G_c \, = \, { (d-2) \, 2^{d-2} \, \pi^{d \over 2} \, \Gamma ( { d \over 2 } ) 
\over \beta_0 } \; \Lambda^{ 2 - d }  \;\; ,
\label{eq:gc_d}
\eeq
and for the mass gap
\beq
m \; = \; \Lambda \, \sqrt{ 2 (d-4) \over 3 (d-2)  \, G_c } \,
\sqrt{ G - G_c } 
\, + \, {\cal O} \left (  G-G_c  \right ) \;\; .
\label{eq:massgap_d}
\eeq
This then leads in $d>4$ to the renormalization group running of Newton's constant
\beq
G ( \mu ) \; = \;  G_c \left [  \, 1 \, + \, 
{ 3 (d-2) \over 2 (d-4)  } \, \left ( { m \over \mu } \right )^2
\, + \, {\cal O} \left (  \left ( { m \over \mu } \right )^4  \right ) \; \right ] \;\; .
\label{eq:grun_d}
\eeq
The dimensionless quantum amplitude is given here in general by $c_0 = 3(d-2)/2(d-4) $ for $d>4$,
and the pole in $d=4$ is then seen as the cause of the log correction seen earlier
in Eq.~(\ref{eq:massgap_4d}).
Conversely, for $d<4 $ the general $d$ result from Eq.~(\ref{eq:gap_eq_d}) is that $\nu = 1 / (d-2) $.
The running of Newton's constant for this case is then given by
\beq
G ( \mu ) \; = \;  G_c \left [  \, 1 \, - \, 
{ (d^2 - 4) \, \pi \, \csc ( \pi d / 2 )  \over 8  } \, 
\left ( { m \over \mu } \right )^{ d-2 }
\, + \, {\cal O} \left (  \left ( { m \over \mu } \right )^{ 2 (d-2) }  \right )  
\right ] \;\; .
\label{eq:grun_da}
\eeq
The previous result of Eq.~(\ref{eq:grun_d}) then implies that for $d > 4 $ 
in real space the quantum correction to Newton's constant always grows 
quadratically with distance as $ \delta G ( r ) \sim ( r / \xi )^2 $, with $\xi = 1/m $
[in agreement with the results of Eqs.~(\ref{eq:grun_4d}) and (\ref{eq:grun_5d})],
while for $d < 4 $ in real space the same quantum correction grows instead
more slowly as $ \delta G ( r ) \sim ( r / \xi )^{d-2} $
[in agreement with the earlier results of Eqs.~(\ref{eq:grun_3d}) and (\ref{eq:grun_2d})].
Figure \ref{fig:exponents} gives a visual comparison of the Regge-Wheeler lattice gravity
vs. Hartree-Fock results for the universal exponent $\nu$ as a function of dimension.
It is noteworthy that the qualitative behavior is similar, although there are some notable
differences, and perhaps, more importantly, the value of $\nu$ in $d=4$.


The previous discussion left largely open the question of what the physical interpretation of
the (self-consistently determined) mass parameter $m$ is.
Recall that it describes a quantity which is zero to all orders in perturbation theory, nevertheless it is
nonzero when an infinite number of diagrams are re-summed, as in the nonlinear gap
equation of Eq.(\ref{eq:gap_eq_d}).
Define the correlation length $\xi = 1/ m$, which on account of its definition as the inverse of
the mass gap and its relationship, via the Lehmann representation of invariant two-point
functions, describes the large distance exponential decay of n-point functions.
The discussion of the previous sections on the nonlinear sigma model in Section \ref{sec:sigma_hf}, 
and on gauge theories in Section \ref{sec:gauge} shows that the correlation length $\xi$ has a
deep relationship with the vacuum condensates of quantum fields.
For a scalar field in the ordered phase in four dimensions one has the non-vanishing 
vacuum condensate (or magnetization for a spin system)
\beq
\langle \, \phi \, \rangle  \;  \simeq \;  { 1 \over \xi } \;\; .
\label{eq:vev_sca}
\eeq
On the other hand, for non-Abelian gauge theories in four dimensions one has the corresponding result, 
describing a non-vanishing vacuum expectation value for the chromo-electric and 
chromo-magnetic fields \cite{cam89,xji95,bro09,dom14} 
\beq
\langle \, F^a_{\mu\nu} \, F^{a \; \mu\nu}  \, \rangle  \;  \simeq \;  { 1 \over \xi^4 } \;\; .
\label{eq:vev_qcd1}
\eeq
In QCD this last result can be obtained from purely dimensional arguments, once the 
existence of a fundamental correlation length $\xi$ (which for QCD is given by the inverse 
mass of the lowest spin zero glueball) is established.
Note that the latter result is completely gauge invariant, in spite of the appearance of the quantity 
$\xi$, which, by virtue of the Hartree-Fock gap equation and its dependence on the gauge-fixed
gluon propagator, could have inherited some gauge dependence.
The power of $\xi$ in the two cases above is thus determined largely by the canonical
dimension of the primary fields $\phi$ or $A_\mu^a $.
The close analogy between non-Abelian gauge theories and gravity then suggests that a similar
identification should be true in gravity for the Ricci scalar, 
\beq
\langle \, g_{\mu\nu} \, R^{\mu\nu}  \, \rangle \; \simeq \;  { 6 \over \xi^2 } \;\; ,
\label{eq:xi_ricci}
\eeq
as argued in \cite{loop07,loop10,vac17}, and reference therein.
Via the classical (or quantum Heisenberg) vacuum field equations $R = 4 \lambda $ one can 
then relate the above quantities to the observed (scaled) cosmological constant 
$\lambda_{\rm obs}$ (up to a numerical constant of proportionality, expected to 
be of order unity).
One obtains
\beq
\third \;  \lambda_{\rm obs} \; \simeq \;  { 1 \over \xi^2 }  \;\; .
\label{eq:xi_lambda}
\eeq 
In this picture the latter quantity is regarded as the quantum gravitational condensate,
a measure of the vacuum energy, and thus of the intrinsic curvature of
the vacuum \cite{loop07,loop10}.
While it is zero to all orders in perturbation theory, it is eventually nonzero as a result of 
nonperturbative graviton condensation, in close analogy to what happens in
non-Abelian gauge theories.

The above considerations can help in providing a quantitative handle 
on the physical magnitude of the new genuinely nonperturbative scale $\xi$.
From the observed value of the cosmological constant 
(see for example the recent 2018 Planck satellite data \cite{pla18}) one
obtains a rough estimate for the absolute magnitude of the scale $\xi$,
\beq
\xi \; \simeq \; \sqrt{3 / \lambda_{\rm obs} } \; \approx \; 5320 \, \text{Mpc} \;\; .
\label{eq:xi_mpc}
\eeq
Irrespective of the specific value of $\xi$, this would indicate
that generally the recovery of classical General Relativity (GR) results takes place at
distances much smaller than the correlation length $\xi$
so that coherent quantum gravity effects become negligible on distance scales $r \ll \xi$.
\footnote{
The actual value for $\lambda_{\rm obs} $, and therefore $\xi$, relies on a multitude of
current cosmological observations, which nowadays is usually analyzed in the 
framework of the standard $\Lambda$CDM model. 
Included in the usual assumptions is the fact that Newton's $G$
does not run with scale. 
If such an assumption were to be relaxed, it could affect 
a number of cosmological parameters, including $\lambda_{\rm obs} $,
whose value could then perhaps change significantly. 
In the following the estimate of Eq.~(\ref{eq:xi_mpc}) is used as a sensible starting point.}
In particular, the static Newtonian potential is expected to acquire a tiny 
quantum correction from the running of $G$ [see Eqs.~(\ref{eq:grun_4d}) and
(\ref{eq:grun_latt})], $ V(r) \sim  - \, G(r) / r $.

\begin{figure}
\begin{center}
       \includegraphics[width=0.7\textwidth]{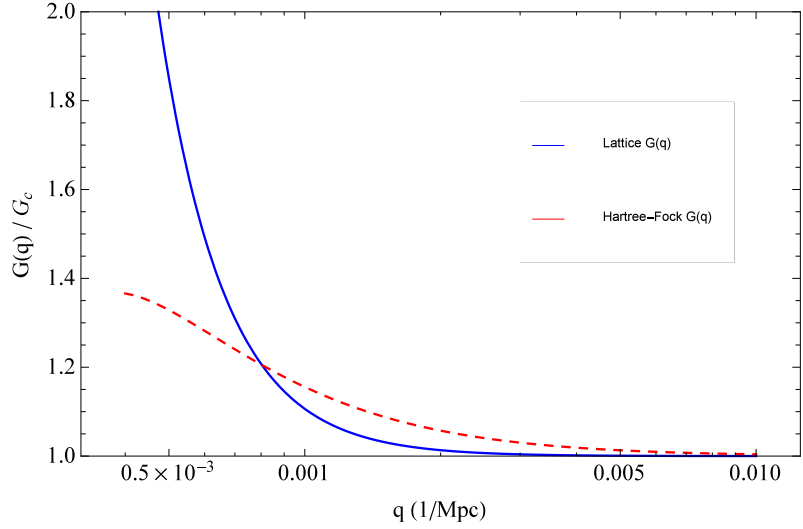}
\end{center}
\caption{
Comparison of Newton's running $G(q)$, as given in Eqs.~(\ref{eq:grun_hf}) (the Hartree-Fock result) 
and (\ref{eq:grun_latt1}) (the Regge-Wheeler lattice result).
The nonperturbative reference scale $\xi$ is directly to the gravitational vacuum condensate, 
as in Eqs.~(\ref{eq:xi_ricci}) and (\ref{eq:xi_lambda}).
}
\label{fig:Gq}
\end{figure}

\begin{figure}
\begin{center}
       \includegraphics[width=0.7\textwidth]{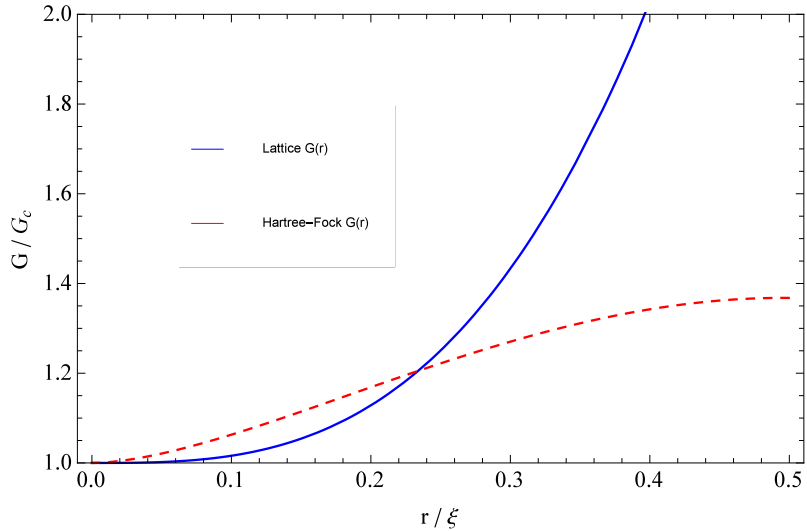}
\end{center}
\caption{
Comparison of the gravitational running coupling $G(r)$ versus $r$, obtained
from $G(q)$ in Eqs.~(\ref{eq:grun_hf}) (the Hartree-Fock result) and (\ref{eq:grun_latt1})
(the Regge-Wheeler lattice result) by setting $q \sim 1/r $.
Note that the approximate Hartree-Fock analytical result of Eq.~(\ref{eq:grun_hf}) (red line) 
initially rises more rapidly for small $r$.
The nonperturbative scale $\xi$ used here is related to the gravitational vacuum condensate
via Eqs.~(\ref{eq:xi_ricci}) and (\ref{eq:xi_lambda}).
}
\label{fig:Gr}
\end{figure}

Figure \ref{fig:Gq} provides a direct comparison between the continuum analytical 
self-consistent Hartee-Fock quantum gravity result for $G(q) $ in Eq.~(\ref{eq:grun_4d}),
\beq
G_{HF} (q) \; = \; G_c \left [ \; 1 \, - \, { 3 \over 2 \, q^2 \, \xi^2 } \, 
\log \left (\, { 3 \over 2 \, q^2 \, \xi^2 } \right ) \, + \, \cdots \; \right ]  \;\; ,
\label{eq:grun_hf} 
\eeq
and the Regge-Wheeler lattice gravity running of $G(q)$ of Eq.~(\ref{eq:grun_latt}) 
as given in \cite{ham15}, 
\beq
G_{Latt} ( q ) \; = \;  G_c \left [  \, 1 \, + \, c_0 \, \left ( { 1 \over q \, \xi } \right )^3
\, + \, {\cal O} \left (  \left ( { 1 \over q \, \xi } \right )^6  \right ) \right ] \;\; .
\label{eq:grun_latt1}
\eeq
Figure \ref{fig:Gr} later provides a direct comparison between the two results, but in real space.
In generating this last graph, we have simply substituted $\mu \rightarrow 1/ r$.
This was done in order to avoid at this early stage the complexities of having to deal, 
in a general coordinate space, with a renormalization group running 
Newton's $ G ( \Box ) $ via the covariant substitution
\beq
-q^2 \; \rightarrow \; \Box  \;\; , \;\;\;\;\;\;\;\;\;\;   G(q) \; \rightarrow \;  G ( \Box )
\label{eq:gbox}
\eeq
with $\Box \equiv  g^{\mu\nu} \nabla_\mu \nabla_\nu $, as discussed
in detail elsewhere, see for example \cite{hw05,rei11}.


To conclude this section, one can raise the legitimate concern of how 
these results are changed by quantum fluctuations of various matter fields
coupled to gravity (scalars, fermions, vector bosons, spin-3/2 fields etc.).
These would enter in the vacuum polarization loop diagrams containing these fields.
Their contribution will appear through many additional loops, and will therefore affect
the value of the coefficient $\beta_0$ appearing in Eq.~(\ref{eq:gap_eq_d}).
Nevertheless one would expect that significant changes to the result of Eq.~(\ref{eq:grun_4d}) 
will arise from matter fields which are light enough to compete with gravity, and 
whose Compton wavelength is therefore comparable to the scale of the 
gravitational vacuum condensate, or equivalently the observed cosmological 
constant, so that for these fields $m^{-1} \sim 1/ \sqrt{\lambda/3}$.
At present the number of candidate fields that could fall into this
category is rather limited, with the photon and a near-massless 
gravitino belonging to this category.
Note that in the $2+\epsilon$ perturbative expansion for quantum gravity
one encounters factors of $25-c$ in the renormalization groups
$\beta$ function, where $c$ is the central charge 
associated with the (massless) matter fields \cite{eps3,eps4}.
In four dimensions similar factors involve $48-c$ \cite{nie81,hug80},
which would again lend support to the argument that such effects should be 
rather small in four dimensions.
This would suggest again that matter loop and radiation corrections are indeed rather small,
unless one has very many massless matter fields.                                                        
Nevertheless, at first one might worry that this would ultimately affect for example the value of 
$\nu$, and thus the power in the running of Newton's constant $G$, as
in Eq.~(\ref{eq:grun_4d}).
But this is not possible, at least within the lowest-order Hartree-Fock approximation
discussed here.
The reason here is the fact that, remarkably, both the power and the amplitude of
the quantum correction to $ G( \mu ) $ in Eqs.~(\ref{eq:grun_4d}), (\ref{eq:grun_d}) and
(\ref{eq:grun_hf}) do not depend on the amplitude $\beta_0$.
The latter only enters in the (non-universal) value of $G_c$.


\vskip 20pt

\section{A Sample Application to Cosmology}
\label{sec:cosmo}

\vskip 10pt

The results presented so far show how the Hartree-Fock approximation applied to quantum 
gravity implies a renormalization group running of Newton's constant, given explicitly in Eq.~(\ref{eq:grun_hf}).
The reference scale $\xi$ appearing in the above expression is given in Eqs.~(\ref{eq:xi_ricci}) 
and (\ref{eq:xi_lambda}), and is seen here as related to the current value of the 
gravitational vacuum condensate, or equivalently to the observed cosmological constant.
In the previous section a comparison of the Hartree-Fock result was made to the numerical Regge-Wheeler lattice
result of Eq.~(\ref{eq:grun_latt1}), which is rather similar in form but nevertheless has both
a larger power and slightly larger amplitude.
In either case, the corrections to a classical (scale-independent) Newton's constant start
to appear only at very large, cosmological distance, on account of the phenomenologically
huge value for $\xi$ in Eq.~(\ref{eq:xi_mpc}).

On cosmological scales, a running of $G$ leads to a multitude of physically observable effects,
ranging from modifications to early cosmological evolution, the primordial growth of matter perturbations, and the cosmic background radiation (CMB).
We chose here to focus on the latter for two main reasons.
The first one is the sensitivity of the CMB data to extremely large scales, clearly comparable in magnitude to the value for $\xi$ given in Eq.~(\ref{eq:xi_mpc}).
The second rather obvious reason is the availability of very accurate recent satellite data presented 
in the Planck15 and Planck18 surveys \cite{pla18}.

Here we will focus almost exclusively on one well-studied physical observable, the matter power spectrum $P_m (k ) $, for which detailed graphs have been produced by the aforementioned 
Planck collaboration.
The matter power spectrum relates to the cosmologically observed matter density correlation
\begin{equation}
G_\rho (r;t,t') 
\equiv 
\left\langle \, \delta_m(\mathbf{x},t) \,\, \delta_m(\mathbf{y},t') \, \right\rangle 
= \frac{1}{V} \int_{V} d^3\mathbf{z} \; \delta_m(\mathbf{x+z},t) \; \delta_m(\mathbf{y+z},t) \;\; ,
\label{eq:delta_rho}
\end{equation}
where $r=\left|\mathbf{x-y}\right|$, and
\begin{equation}
\delta_m (\mathbf{x},t) \equiv \frac{\delta\rho(\mathbf{x},t)}{\bar{\rho}(t)} 
= \frac{\rho(\mathbf{x},t) - \bar{\rho} (t) }{\bar{\rho}(t)} \; \; .
\label{eq:delta_def}
\end{equation}
is the matter density contrast, which measures the fractional overdensity of matter denstiy $\rho$ above an average background density $\bar{\rho}$.
In the literature, this correlation is more often studied in wavenumber-space, 
$G_\rho ( \mathbf{k} ;t,t') \equiv \left\langle \, \delta ( \mathbf{k} ,t) \, \delta (- \mathbf{k} ,t') \, \right\rangle$, 
via a Fourier transform. 
It is also common to bring these measurements to the same time, 
say $t_0$, so that one can compare density fluctuations of different scales 
as they are measured and appear today.  
The resultant object $P_m(k)$ is referred to as the matter power spectrum,
\begin{equation}
P_m (k) \equiv 
(2\pi)^3 \langle \; \left| \delta ( \mathbf{k} ,t_0) \right|^2 \; \rangle =
(2\pi)^3  F(t_0)^2   \langle \; \left| \Delta ( \mathbf{k} ,t_0) \right|^2 \; \rangle \;\; ,
\label{eq:pm_def}
\end{equation}
where $\delta( \mathbf{k} ,t) \equiv F(t) \, \Delta( \mathbf{k} ,t_0)$.  
The factor $F(t)$ then simply follows the standard General Relativity evolution formulas, as governed 
by the Friedmann-Robertson-Walker (FRW) metric.  
As a result, $P_m(k)$ can be related to, and extracted from, the real-space measurements 
via the inverse transform
\bea
G_\rho (r;t,t') 
& = &
\int \frac{d^3 k}{(2\pi)^3} \; G_\rho (k;t,t') \; 
e^{-i \mathbf{k} \cdot (\mathbf{x}-\mathbf{y})} \\
& = &
\frac{1}{2 \pi^2} \frac{F(t)F(t')}{F(t_0)^2} \; 
\int_{0}^{\infty} dk \; k^2 \; P_m (k) \; \frac{\sin{(kr)}}{kr}  \;\;  .
\label{eq:pm_k}
\eea
It is common to parameterize these correlators by a so-called scale-invariant spectrum, 
which includes an amplitude and a scaling index, conventionally written as 
\begin{equation}
G_\rho (r;t_0,t_0 ) = \left( \frac{r_0}{r} \right) ^ \gamma \;\; ,
\label{eq:gamma_def}
\end{equation}
or equivalently 
\begin{equation}
P_m (k) = \frac{a_0}{k^s}  \;\;  ,
\label{eq:s_def}
\end{equation}
with the powers related by $s = (d-1) - \gamma = 3 - \gamma $ via the Fourier transform.  
Nevertheless, such simple parametrizations generally only apply to specific regions
in wave-vector space (specifically the galaxy regime at larger $k$, or the primordial CMB regime
at smaller $k$), in spite of the fact that different regimes are known to be
related to each other by standard classical cosmological evolution, via the so-called transfer function.

In the matter-dominated era, such as the one where galaxies and clusters are formed, the energy momentum tensor follows a perfect pressureless fluid to first approximation.  
Hence, the trace equation reads
\begin{equation}
R - 4 \lambda = - 8 \pi G \, T  \;\; .
\label{eq:trace}
\end{equation}
(For any perfect fluid the trace gives $T = 3 p - \rho$, and thus 
$ T \simeq - \rho $ for a non-relativistic fluid.)  
Since $\lambda$ is a constant, the curvature and matter variations, and hence their correlations, 
are directly related
\begin{equation}
\langle \; \delta R (x) \, \delta R (y) \; \rangle \, = \, 
(8 \pi G)^2 \; \langle \; \delta \rho (x) \, \delta \rho (y) \; \rangle  \;\; .
\label{eq:delta_r}
\end{equation}
A number of subtleties arise here, such as the fact that 
to extend beyond the linear matter dominated regime, the trace equation alone becomes insufficient (since the trace of the energy momentum tensor for radiation vanishes), 
and the full tensor field equations have to be used.  
Furthermore, in a real universe with multiple fluid components, interactions and transient behaviors have to be taken into account, which are governed by coupled nonlinear Boltzmann equations.  
However, these classical procedures have been fully worked out in standard 
cosmology texts \cite{wei08,dod03}, as well as in currently popular sophisticated computer codes,
such as CAMB and MGCAMB \cite{lew99,lew20,zuc19}, CLASS and MGCLASS \cite{les11,tes15},
ISiTGR \cite{gar19,gar20,lew02} and COSMOMC \cite{lew02} (for a detailed comparison of features 
in various codes see \cite{cod18}).

Quite generally, the matter power spectrum can be decomposed as two parts -- an initial condition known as the primordial spectrum $\mathcal{R}^o_k$, and an interpolating function between the domains known as a transfer function $\mathcal{T}(k)$ \cite{wei08}. 
As a result, the full matter power spectrum $P_m (k)$ beyond the galaxy (larger $k$) domain will 
take the form
\begin{equation}
P_{m} (k) \, = \, C_0 \, 
\left( \mathcal{R}^o_k \right)^2 k^4 \, \left[ \mathcal{T}(\kappa) \right]^2  \;\;\;  ,
\label{eq:Pkfull}
\end{equation}
where $C_0  \equiv 4 (2 \pi )^2 \, C^2 ( \Omega_\Lambda / \Omega_M ) / 25 \, 
\Omega_M^2 H_0^4 $ is a combined constant of cosmological parameters, and the $k^4$ factor is
purely for convenience.  
The transfer function is usually written in terms of $\kappa\equiv \sqrt{2}k/k_{eq}$, a scaled dimensionless wavenumber, with $k_{eq}$ being the wavenumber at matter-radiation-equality.  
With this decomposition, the transfer function is a fully classical  solution of the set of Friedmann and Boltzmann equations, capturing the nonlinear dynamics.  
This leaves the initial primordial function, which is usually parameterized as a scale-invariant spectrum 
\begin{equation}
( \mathcal{R}_k^0 )^2 \; = \;
N^2 \, \frac{1}{k^3} \, \left( \frac{k}{ k_{ \mathcal{R}} } \right)^{n_s - 1}  \;\; ,
\label{eq:Rksq_param}
\end{equation}
involving an overall amplitude $N^2$ and a spectral index $n_s$.  
$ k_{ \mathcal{R} } $ is commonly referred to as the ``pivot scale'', a reference scale conventionally taken to be $ k_{ \mathcal{R} } = 0.05 \, \rm{Mpc}^{-1}$ in cosmology.  
While the transfer function $\mathcal{T} (\kappa)$ -- the solution to the coupled and nonlinear set of Friedmann, Boltzmann and continuity differential equations -- is difficult to obtain as an explicit function, it is nevertheless in principle fully determined from classical dynamics.  
Moreover, assuming standard $\Lambda$CDM cosmology dynamics and evolution, a semi-analytical interpolating formula for $\mathcal{T}(\kappa)$ \cite{wei08} is known.  
As a result, if the initial spectrum $\mathcal{R}^o_k$, or more specifically the parameters $N$ and $n_s$, are fixed, then $P_m(k)$ is fully determined.  

Here, in this section, the main concern will be the inclusion of the effects of a running of Newton's 
$G$, as given in Eqs.~(\ref{eq:grun_hf}), (\ref{eq:grun_latt1}) and (\ref{eq:xi_lambda}) in the matter power spectrum, as measured for example in current CMB data.
Modifications to the matter power spectrum $P_m (k)$ at small $k$ originating in a running $G(k)$ 
can be done either analytically using the transfer function \cite{hyu18,hyu19} or by relying on more comprehensive numerical programs \cite{hyp20}.  
Analytically, the effect of a running Newton's constant [Eq.~(\ref{eq:grun_hf})] can be included via dimensional analysis for the correct factors of $G$ to include, $ P_m (k) \,\rightarrow \, ( G_0 / G(k) )^2 P_m (k)  $ \cite{hyu19}.  
Here we will focus instead on the more comprehensive, and more accurate, results that can be obtained
by consistently incorporating the running of $G$ in all relevant cosmological equations, 
and in many more cosmological observables, such as temperature and polarization correlations \cite{hyp20}.

\begin{figure}
\begin{center}
       \includegraphics[width=0.90\textwidth]{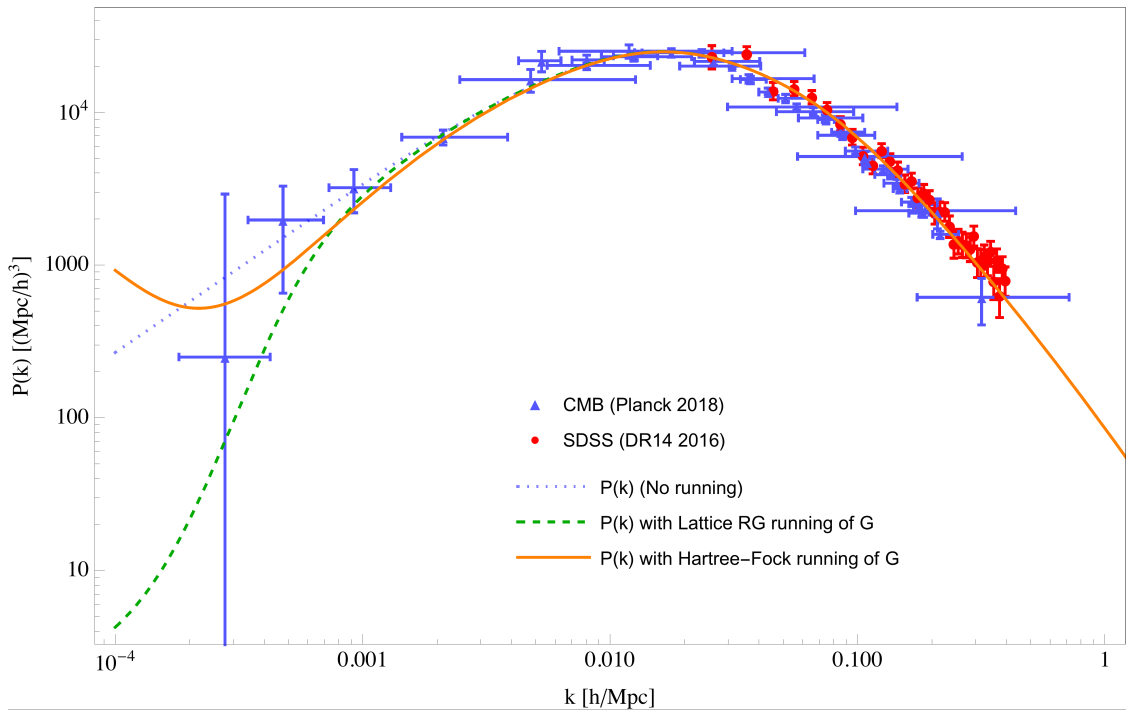}
\end{center}
\caption{Matter power spectrum $P(k)$ with lattice versus Hartree-Fock running of Newton's constant $ G $. The middle solid (orange) curve shows $ P(k) $ implemented with the Hartree-Fock 
running of Newton's constant $ G $ factor as given in Eq.~(\ref{eq:grun_hf}).
The lower dashed (green) curve shows the original lattice RG running of 
Newton's constant $ G $ (with the original lattice 
coefficient of Eq.~(\ref{eq:grun_latt1}) equal to $ c_0 = 16.0 $) for comparison.  
The original spectrum with no running is also displayed by the top 
dotted (blue) curve for reference.}
\label{fig:pkrun_hf}
\end{figure}

The key resultant prediction for the matter power spectrum $P_m (k)$ is found here in 
Figure \ref{fig:pkrun_hf}, showing (not surprisingly) still an almost perfect fit to all observational 
data for $k \gg m \simeq 2.8\times 10^{-4} \, h\text{Mpc}^{-1}$.
Nevertheless, for scales of $k$ comparable to $m$, additional quantum effects
become significant due to the $G(k)$, enough to cause significant deviations from the classical 
$\Lambda$CDM result for $P_m (k)$. 
In Figure \ref{fig:pkrun_hf}, the middle solid orange curve shows the Hartree-Fock expression for the 
running of Newton's constant $ G $, while the bottom dashed green curve and the top blue dotted 
curve show the lattice result of a running Newton's constant $ G $ 
(with the lattice coefficient $ c_0=16.0 $), as well as the result for no running respectively 
for reference.  
It seems that the Hartree-Fock running of $G$ is in reasonably good consistency with the lattice expression, except for the eventual unwieldly upturn below $ k < 2 \times 10^{-4} \, h \text{Mpc}^{-1}$.  
However, this upturn is most likely an artifact from the Hartree-Fock expression being 
just a first-order analytical approximation after all (as discussed earlier, the lowest order
Hartree-Fock approximation can be extended to higher order, by including increasingly 
complex higher loop diagrams, with dressed propagators and vertices still determined 
self-consistently by a truncated version of the Schwinger-Dyson equations).
Nevertheless, the Hartree-Fock approximation shows good consistency with both 
the latest available observational  data sets, as well as with the lattice result.  
The fact that it exhibits a gentler dip at small $k$ perhaps also provides support for 
a potentially smaller lattice running coefficient of Eq.~(\ref{eq:grun_latt1}) of approximate value 
$ c_0 \approx 2.29 $.  
More details for the most recent constraints on this lattice parameter $c_0$ are discussed 
in \cite{hyu19,hyp20}.  


Following similar analysis to determining $P_m(k)$, other spectra such as the angular temperature spectrum $C_l^{TT}$, should be fully derivable from the primordial function $\mathcal{R}^o_k$, or specifically $n_s$, which is set by the scaling of gravitational curvature fluctuations $\nu$.


\vskip 20pt

\section{Conclusion}
\label{sec:conclusion}

\vskip 10pt

In this work we have shown how the Hartree-Fock approximation to quantum gravity 
can be carried out, by considering what ultimately reduces to a set of
relatively easy to evaluate one-loop integrals.
The main feature of such an approach is that the complete analytic expression for the 
one-loop integrals allows one to write down a gap equation for the dynamically generated
mass scale in the strongly coupled phase and, more importantly, for any spacetime dimension.
This result can be viewed as analogous to the situation in the nonlinear sigma model,
where a nonlinear gap equation is obtained explicitly, and can then in principle be solved for 
any dimension.
Due to the inherent approximation of a mean-field type approach the calculation is 
expected to be rather crude, but nevertheless provides interesting insights, and can
furthermore provide a useful starting point for improved, higher loop self-consistent
calculations.

There are a number of interesting features that arise from the results presented earlier,
which will be summarized in the following paragraphs.
The first notable aspect is the appearance of a nontrivial renormalization group
fixed point (referred to as a critical point in statistical field theory language) $G_c$ in Newton's constant,
for any spacetime dimension $d$ greater than two.
As a consequence, above two dimensions the theory exhibits two phases, a Coulomb-like
phase with gravitational screening for $ G < G_c $, and a strongly coupled phase with 
gravitational antiscreening for $ G > G_c $.
Since the Hartree-Fock method retains some vestiges of a perturbative diagrammatic 
calculation, the above results do not allow one to determine whether both phases 
are indeed physical. 
On the other hand current lattice calculations, which are genuinely nonperturbative, suggest
that gravitational screening is impossible (no stable ground state exists $G<G_c$) 
and that, as in Yang Mills theories, only the gravitational antiscreening 
phase for $ G > G_c$ is physically realized in this theory.

For the latter phase the Hartree-Fock calculation gives, once suitable renormalization group 
arguments are applied, an explicit expression for the running of the 
gravitational Newton's constant as a function of scale.
The latter can be expressed as a $ G(k) $ in wavevector space, or, if one wishes, 
more generally as a covariantly formulated $ G (\Box)$, with covariant d'Alembertian
$\Box \equiv  g^{\mu\nu} \nabla_\mu \nabla_\nu $.
Unlike perturbatively renormalizable theories such as QCD, the RG running of $G$ in the 
gravitational case is not logarithmic, but instead follows a power law in the relevant scale.
In addition, the Hartree-Fock results suggest that for quantum gravity the lower critical dimension is two
(as expected based on naive dimensional arguments, and also on the basis of the lattice results,
and initially from Wilson's $2+\epsilon$ gravity perturbative expansion as well), and furthermore 
that the upper critical dimension is six. 
This would correlate with the fact that, within the Hartree-Fock approximation, above six 
spacetime dimensions scaling dimensions and critical exponents flow into their 
Gaussian, free field value. 
How that would explicitly reflect on the nature of various (local and non-local)
gravitational correlations is still not entirely clear at this point.

More importantly, in the physically relevant case of four spacetime dimensions 
one finds that the gravitational coupling grows like a distance squared, up to 
logarithmic corrections (between $d=2$ and $d=4$ the relevant power varies as $1 / (d-2) $, 
and stays constant above that). 
This result is similar to, but nevertheless still quantitatively different from, the Regge-Wheeler 
lattice result, which gives a renormalization group growth of the gravitational 
coupling $G$ with the cube of the distance, see \cite{ham15,vac17} and references therein.
Nevertheless, in either case the reference scale for the growth of Newton's constant as 
a function of distance is determined by a new, genuinely nonperturbative quantity
$ \xi = 1 / m $.
The latter is non-analytic in the bare couplings, and arises naturally as an 
integration constant of the renormalization group equations.
It was argued elsewhere \cite{loop07,vac17} that this new nonperturbative scale, specific to 
quantum gravity, should be identified with the gravitational vacuum condensate 
$\langle R \rangle = 2 \lambda $ via $ \lambda = 3 / \xi^2 $, where $\lambda$ is 
the observed (tiny) scaled cosmological constant, in this picture more properly referred to
as the gravitational vacuum energy.
The latter would then provide the needed reference scale for the RG running of the 
gravitational coupling $G$.
Since $\xi $ is very large (from current astrophysical observations of $\lambda$ 
\cite{pla18}) one has $\xi \sim 5300 {\rm Mpc}$),
which would then lead to a very slow rise of $G$, observable only on very large, 
cosmological scales.

It is of some interest here to point out that the Hartree-Fock approximate calculation 
does not just give values for exponents and scaling dimensions, it also provides an 
explicit analytic expression for quantum amplitudes.
Such as, for example, the overall amplitude of the quantum correction to $G (k)$
when referred to the new nonperturbative scale $\xi$.
This is quite different from ordinary perturbation theory say in the $2 + \epsilon $ expansion,
where no information can ever be gained on the quantum amplitudes themselves
(only on some ratios of critical amplitudes, as is already the case in the nonlinear sigma model).

What is the physical interpretation of the self-consistent Hartree-Fock type calculation?
For suggestions, one can look at the analogous results in condensed matter physics,
or from the nonlinear sigma model,
to try to gain some insight into what type of physical process is taken into account in the 
Hartree-Fock self-consistent, mean field-type approximation.
Perturbation theory initially only takes into account single graviton exchange (at the tree level), 
then to higher order two graviton exchanges in loop diagrams, then three graviton exchanges etc..
On the other hand, in the Hartree-Fock approximation all these multi-graviton effects are included into 
one single, self-consistently determined, diagrammatic contribution.
The latter sums up infinitely many perturbative diagrams, and thus can lead to entirely 
novel features such as a dynamically generated nonperturbative scale. 
The lowest order Hartree-Fock self-consistent approximation should therefore be viewed as a 
(physically motivated) re-summation of a certain subset of infinitely many diagrams.
The next order Hartree-Fock correction then self-consistently re-sums a second set of topologically 
distinct more complex diagrams, leading presumably to a further improvement over the 
lowest order Hartree-Fock result.
It is therefore noteworthy here that the emergence of the new nonperturbative scale 
$\xi$ can only arise if infinitely many graviton loop diagrams are resummed.
The various loop diagrams then all add up coherently, and give rise to the new phenomenon 
of a nonvanishing gravitational vacuum condensate.
Nevertheless, in the Hartree-Fock approximation the connection between the nonperturbative mass
scale $m=1/\xi$ and the gravitational field vacuum condensate $ \langle R \rangle $ is not immediate.
One way to justify such as result is to consider the fact that the renormalized, effective 
cosmological $\lambda$ constant (related to the vacuum expectation value of the scalar curvature
via the Heisenberg equations of motion) clearly appears as a mass-like term in the weak field 
expansion of the gravitational action.
Nevertheless, in the gravity case one has the well-known subtlety that such a mass-like 
term remains fully consistent with general covariance (and thus entirely independent of gauge
choice), since the $\lambda$ term in the gravitational action retains that property.

\vspace{30pt}

\newpage




\vfill


\end{document}